\documentclass[useAMS,usenatbib]{mn2e}
\usepackage{amssymb}
\usepackage{amsmath}
\usepackage{times}
\usepackage{graphicx}
\usepackage{wasysym}

\newcommand{\mincir}{\raise
-2.truept\hbox{\rlap{\hbox{$\sim$}}\raise5.truept 
\hbox{$<$}\ }}
\newcommand{\magcir}{\raise
-2.truept\hbox{\rlap{\hbox{$\sim$}}\raise5.truept
\hbox{$>$}\ }}
\newcommand{\minmag}{\raise-2.truept\hbox{\rlap{\hbox{$<$}}\raise
6.truept\hbox
{$>$}\ }}

\newcommand{\be}{\begin{equation}}
\newcommand{\ee}{\end{equation}}
\newcommand{\ba}{\begin{eqnarray}}
\newcommand{\ea}{\end{eqnarray}}
\newcommand{\brr}{\begin{array}}
 
\newcommand{\err}{\end{array}}
\newcommand{\bc}{\begin{center}}
\newcommand{\ec}{\end{center}}

\newcommand{\tu}{\textunderscore}

\DeclareMathAlphabet{\mathsc}{OT1}{cmr}{m}{sc}
\def\testbx{bx}%
\DeclareRobustCommand{\ion}[2]{%
\relax\ifmmode
\ifx\testbx\f@series
{\mathbf{#1\,\mathsc{#2}}}\else
{\mathrm{#1\,\mathsc{#2}}}\fi
\else\textup{#1\,{\mdseries\textsc{#2}}}%
\fi}

\title[GSMF and SFR$-M_{\rm \star}$ relations of high-$z$ galaxies]{The stellar mass function and star formation rate${\bf{-}}$stellar mass relation of galaxies at ${\bf{z\sim4-7}}$}

 \author[A. Katsianis et al.]{A. Katsianis$^{1,2}$\thanks{E-mail:
     kata@student.unimelb.edu.au }, E. Tescari$^{1,2}$ and J. S. B. Wyithe$^{1,2}$\\ \\ $^1$ School of Physics, The University of Melbourne, Parkville, VIC 3010, Australia \\ $^2$ ARC Centre of Excellence for All-Sky Astrophysics (CAASTRO) \\}

\begin{document}

\maketitle

\begin{abstract}

We investigate the evolution of the star formation rate$-$stellar mass relation (SFR$-{\rm M}_{\star}$) and Galaxy Stellar Mass Function (GSMF) of $z\sim 4-7 $ galaxies, using cosmological simulations run with the smoothed particle hydrodynamics code {\small{P-GADGET3(XXL)}}. We explore the effects of different feedback prescriptions (supernova driven galactic winds and AGN feedback), initial stellar mass functions and metal cooling. We show that our fiducial model, with strong energy-driven winds and early AGN feedback, is able to reproduce the observed stellar mass function obtained from Lyman-break selected samples of star forming galaxies at redshift $6\le z\le7$. At $z\sim4$, observed estimates of the GSMF vary according to how the sample was selected. Our simulations are more consistent with recent results from  K-selected samples, which provide a better proxy of stellar masses and are more complete at the high mass end of the distribution. We find that in some cases simulated and observed SFR$-{\rm M}_{\star}$ relations are in tension, and this can lead to numerical predictions for the GSMF in excess of the GSMF observed. By combining the simulated SFR(M$_{\star}$) relationship with the observed star formation rate function at a given redshift, we argue that this disagreement may be the result of the uncertainty in the SFR$-{\rm M}_{\star}$ ($L_{\rm UV}-{\rm M}_{\star}$) conversion. Our simulations predict a population of faint galaxies not seen by current observations.

\end{abstract}


\section{Introduction}

Our knowledge of galaxies in the early Universe, has expanded substantially over the last ten years. Galaxies are identified out to $z \sim 8$ and beyond, and their key quantities like Star Formation Rate (SFR), stellar mass (M$_{\star}$), dust extinction and age can be measured via spectroscopy and multi-wavelength photometry \citep{Finkelstein11,Lee2012,Dunlop12,bouwens2012,Gonzalez12}. The Star Formation History (SFH) is thought to be determined by a combination of the formation and growth of Dark Matter (DM) halos, the stellar Initial Mass Function (IMF), and astrophysical processes such as gas accretion, stellar mass loss, radiative cooling, and feedback from Supernovae (SN) and Active Galactic Nuclei (AGN). The SFH of high redshift galaxies is characterised by a rapid rise of the star formation rate until redshift $z\sim6$. Then, there is a period of more slowly increasing star formation down to $z\sim2$. Finally, the SFR has been found to decrease by a factor of 30 from $z\sim2$ to $z = 0$ \citep{Daddi2007}. The relation between SFR and galaxies is also probed by the specific SFR (sSFR = SFR$/$M$_{\star}$). \citet{Gonzalez10} found a roughly constant sSFR from $z\sim7$ to $z\sim2$. More recently \citet{Smit2013} and \citet{StarkSchen} provided evidence that the sSFR increases past $z \sim 2$ once nebular emission lines are accounted for.

The cosmic star formation rate density and Star Formation Rate Functions (SFRFs) can be useful for constraining theoretical models, since they provide a good physical description of galactic growth over time \citep{smit12,Vogel13}. For example, \citet{TescariKaW2013} used hydrodynamic simulations to model the observed SFRFs of \citet{smit12}. Another key measurement that constraints the star formation history, is the location of galaxies on the SFR$-{\rm M}_{\star}$ plane \citep{Lee2012}. There is strong evidence of a correlation between SFR and stellar mass at all redshifts, from $z=0$ to the earliest observed epochs at $z\sim7$ \citep{stark2009,Labbe10,Magdis10,Gonzalez11,Whitaker2012}. This relation is observed to have little apparent evolution between $z\sim4$ and $z\sim7$ \citep{Gonzalez11,MClure2011,Gonzalez12} unlike predictions from cosmological simulations and theoretical models \citep{Weinmann2011,Wilkins13}. 

 A number of authors \citep[e.g.][]{Dave08,Dutton10,Finlator11,DayalFerrara2012,Haas2013,Kannan2014} have used hydrodynamic and semi-analytic models to predict the SFR$-{\rm M}_{\star}$ relation for low and intermediate redshifts. \citet{Dave08} studied the SFR$-{\rm M}_{\star}$ relation at $z\le 2$ using hydrodynamic simulations. At $z=0$, the simulated SFR$-{\rm M}_{\star}$ relation is generally in agreement with observations, though the observed slope of the relation is somewhat shallower than predicted. By $z = 1$, the slopes predicted by hydrodynamic simulations are similar to those observed by \citet{Elbaz07} but steeper than those of \citet{Noeske2007}. At $z = 2$ numerical results also predict a relation that is steeper than found in the observations of \citet{Daddi2007}. \citet{Dutton10} used a semi-analytic model for disk galaxies to explore the origin of the time evolution and scatter of the SFR$-{\rm M}_{\star}$ relation at $z=0-6$. As with hydrodynamic results, the simulated relation is generally in agreement with observations, although the observed slope of the relation is shallower than predicted. At $z\sim2$, the semi-analytic model underpredicts the SFR at a fixed mass compared with the observations of \citet{Daddi2007}, and is in agreement with the simulations of \citet{Dave08}. Moreover, the semi-analytic results of \citet{Dutton10} are consistent with the observations of \citet{Daddi2009} at $z\sim4$, but at $z\sim6$ there is a tension with the results of \citet{Gonzalez10} (i.e. the predicted SFR at a fixed mass is larger than observed).

 A key challenge facing  models of galaxy formation is to explain why the shape of the dark matter halo mass function and the Galaxy Stellar Mass Function (GSMF) are different. The important factor that affects the shape of the GSMF and explains the differences of the two distributions is generally thought to be feedback. Cosmological hydrodynamic simulations that take into account radiative cooling, but do not implement any feedback mechanisms linked with star formation, overpredict the stellar mass within halos \citep{Balogh01}. Moreover, the observed slope of the GSMF is substantially flatter than the GSMF obtained from hydrodynamic simulations \citep{choina11}. \citet{PuchweinSpri12} used {\small{GADGET-3}} simulations to investigate the impact of SN driven galactic winds and AGN feedback on the shape of stellar mass function at $z\le2$. By adopting a scheme where wind velocities are proportional to the escape velocity of each galaxy \citep{Martin05}, they were able to reproduce the low mass end of the observed GSMF. On the other hand, AGN feedback is crucial to shape the high mass end of the GSMF at low redshift.

 Galaxy stellar mass functions at high redshift ($z>4$) are very difficult to measure directly due to selection effects but can also provide constraints on scenarios of galaxy formation and early evolution \citep{Marchesini09,Caputi11,Gonzalez11,Lee2012,Santini12}. For $4\le z\le7$, the available estimates of the GSMF are based mostly on UV-selected samples that are incomplete in mass and/or are usually derived by adopting median or average mass-to-light ratios for all galaxies, rather than detailed object-by-object estimates. In the case of  K-selected samples, surveys are more complete but they are limited to the most massive objects at these redshifts.

 This paper is the second of a series in which we present the results of the {\textsc{Angus}} ({\textit{AustraliaN {\small{GADGET-3}} early Universe Simulations}}) project.  In the first paper of the series \citep{TescariKaW2013} we constrained and compared our hydrodynamic simulations with observations of the cosmic star formation rate density and SFRF. We showed that a fiducial model with strong-energy driven winds and early AGN feedback is needed to obtain the observed SFRF of galaxies at $z \sim4-7$ \citep{smit12}. In this work we investigate the driving mechanisms for the evolution of the GSMF and SFR$-{\rm M}_{\star}$ relation at redshift $4\le z\le 7$ using the same set of cosmological simulations as in \citet{TescariKaW2013}. We explore different feedback prescriptions, in order to understand the origin of the difference between observed and simulated relationships. We also study the effects of metal cooling and choice of IMF. We do not explore the broad possible range of simulations, but concentrate on the simulations that can describe the high-$z$ SFRF function.

This paper is organized as follows. In Section \ref{sec_sim} we present a brief description of our simulations along with the different feedback models used.  In Section \ref{obdatasampSFRSMall} we study the star formation rate in halos. In particular, in Section \ref{obdatasampSFRSM} we present the observed SFR$-{\rm M}_{\star}$ relations. In Section \ref{CompaSFRSM} we compare our simulated results with observations. In section \ref{Mass function} we discuss observed GSMFs from four different sets of data and compare with our simulated results. In section \ref{bmdis} we discuss our best fiducial model. Finally, in Section \ref{concl} we summarise our main results and present our conclusions.

\section{The simulations}
\label{sec_sim}

In this work we use the set of {\textit{AustraliaN {\small{GADGET-3}} early Universe Simulations}} ({\textsc{Angus}}) described in \citet{TescariKaW2013}. We run these simulations using the hydrodynamic code {\small{P-GADGET3(XXL)}}\footnote{The features of our code are extensively described in \citet{TescariKaW2013}, therefore we refer the reader to that paper for additional information.}, a modified version of {\small{GADGET-3}} \citep[last described in][]{springel2005}.

We assume a flat $\Lambda$ cold dark matter ($\Lambda$CDM) model with\footnote{This set of cosmological parameters is the combination of 7-year data from WMAP \citep{komatsu11} with the distance measurements from the baryon acoustic oscillations in the distribution of galaxies \citep{percival10} and the Hubble constant measurement of \citet{riess09}. Note that some of these
  parameters are in tension with recent results from the
  \textit{Planck} satellite \citep{Planck13}.} $\Omega_{\rm 0m}=0.272$, $\Omega_{\rm 0b}=0.0456$, $\Omega_{\rm \Lambda}=0.728$, $n_{\rm
  s}=0.963$, $H_{\rm 0}=70.4$ km s$^{-1}$ Mpc$^{-1}$ (i.e. $h = 0.704$) and $\sigma_{\rm 8}=0.809$. The simulations cover a cosmological volume with periodic boundary conditions initially occupied by an equal number of gas and dark matter particles. We adopt the multiphase star formation criterion of \citet{springel2003} in which the Inter Stellar Medium (ISM) changes phases under the effect of star formation, evaporation, restoration and cooling. With this prescription baryonic matter is in the form of a hot or cold phase, or in stars. In this model, whenever a gas particle reaches a density larger than a given threshold density $\rho_{\rm th}$, it is considered to be star-forming. A typical value for $\rho_{\rm th}$ is $\sim0.1$ cm$^{-3}$ (in terms of the number density of hydrogen atoms), but the exact density threshold is calculated according to the IMF used and the inclusion/exclusion of metal-line cooling.

Our code follows the evolution of 11 elements (H, He, C, Ca, O, Ne, Mg, S, Si and Fe) released from supernovae (SNIa and SNII) and low and intermediate mass stars self-consistently \citep{T07}. Radiative cooling and heating processes are included by following the procedure presented in \citet{wiersma09}. We assume a mean background radiation composed of the cosmic microwave background and the \citet{haardtmadau01} ultraviolet/X-ray background from quasars and galaxies. Contributions to cooling from each one of the eleven elements mentioned above have been pre–computed using the {\small {Cloudy}} photo–ionisation code \citep[last described in][]{ferland13} for an optically thin gas in (photo)ionisation equilibrium. In this work we use cooling tables for gas of primordial composition (H + He) as the reference configuration. To test the effect of metal-line cooling, we include it in one simulation.

\begin{table*}
\centering
\begin{tabular}{llccccccc}
  \\ \hline & Run & IMF & Box Size & N$_{\rm TOT}$ & M$_{\rm DM}$ & M$_{\rm GAS}$  &
  Comoving Softening & Feedback \\ & & & [Mpc/$h$] &
  & [M$_{\rm \odot}$/$h$] & [M$_{\rm \odot}$/$h$] & [kpc/$h$] \\ \hline
  & \textit{Kr24\tu eA\tu sW} & Kroupa & 24 & $2\times288^3$ &
  3.64$\times10^{7}$ & $7.32\times 10^6$ & 4.0 & Early AGN $+$ Strong Winds \\
  & \textit{Ch24\tu lA\tu wW} & Chabrier & 24 & $2\times288^3$ &
  3.64$\times10^{7}$ & $7.32\times 10^6$ & 4.0 & Late AGN $+$ Weak Winds \\
  & \textit{Sa24\tu eA\tu wW} & Salpeter & 24 & $2\times288^3$ &
  3.64$\times10^{7}$ & $7.32\times 10^6$ & 4.0 & Early AGN $+$ Weak Winds \\
  & \textit{Ch24\tu eA\tu sW} & Chabrier & 24 & $2\times288^3$ & 3.64$\times10^{7}$ & $7.32\times 10^6$ & 4.0 & Early AGN $+$ Strong Winds  \\
  & \textit{Ch24\tu lA\tu sW} & Chabrier & 24 & $2\times288^3$ &
  3.64$\times10^{7}$ & $7.32\times 10^6$ & 4.0 & Late AGN $+$ Strong Winds \\
  & \textit{Ch24\tu eA\tu vsW} & Chabrier & 24 & $2\times288^3$ &
  3.64$\times10^{7}$ & $7.32\times 10^6$ & 4.0 & Early AGN $+$ Very Strong Winds \\
  & \textit{Ch24\tu NF} & Chabrier & 24 & $2\times288^3$ &
  3.64$\times10^{7}$ & $7.32\times 10^6$ & 4.0 & No Feedback\\
  & \textit{Ch24\tu Zc\tu eA\tu sW}$^a$ & Chabrier & 24 &
  $2\times288^3$ & 3.64$\times10^{7}$ & 
  $7.32\times 10^6$ & 4.0 & Early AGN $+$ Strong Winds \\
  & \textit{Ch24\tu eA\tu MDW}$^b$ & Chabrier & 24 & $2\times288^3$ &
  3.64$\times10^{7}$ & $7.32\times 10^6$ & 4.0 & Early AGN $+$ \\
  & & & & & & & & Momentum-Driven Winds \\
  & \textit{Ch18\tu lA\tu wW} & Chabrier & 18 & $2\times384^3$ &
  6.47$\times10^{6}$ & $1.30\times 10^6$ & 2.0 & Late AGN $+$ Weak Winds \\
  & \textit{Ch12\tu eA\tu sW} & Chabrier & 12 & $2\times384^3$ &
  1.92$\times10^{6}$ & $3.86\times 10^5$ & 1.5 & Early AGN $+$ Strong Winds \\
  \hline \\
\end{tabular}
\caption{Summary of the different runs used in this work. Column 1, run name; column 2,
  Initial Mass Function (IMF) chosen; column 3, box size in comoving Mpc/$h$;
  column 4, total number of particles (N$_{\rm TOT} =$ N$_{\rm
    GAS}$ $+$ N$_{\rm DM}$); column 5, mass of the dark matter particles; column 6, initial mass of the gas particles;
  column 7, Plummer-equivalent comoving gravitational softening length; column 8, type of feedback
  implemented. See Section \ref{simout} for
  more details on the parameters used for the different feedback
  recipes. $(a)$: in this simulation the effect of metal-line cooling is
  included. $(b)$: in this simulation we
  adopt variable momentum-driven galactic winds. In all the other simulations
  galactic winds are energy-driven and the wind velocity is constant (Section \ref{subSup}).}
\label{tab:sim_runs}
\end{table*}

A range of initial stellar mass functions can be employed. For this work we used three different IMFs:
\begin{itemize}
\item \citet{salpeter55}:
\begin{eqnarray}
  \xi(m)=0.172\times m^{-1.35}
\end{eqnarray}
\item \citet{kroupa93}:
\begin{eqnarray}
  \xi(m)=\left\{\begin{array}{l}0.579\times m^{-0.3}\ \ 0.1\,{\rm M}_{\rm
        \odot}\le m <0.5\,{\rm M}_{\rm
        \odot}\\0.310\times m^{-1.2} \ \ 0.5\,{\rm M}_{\rm
        \odot}\le m<1\,{\rm M}_{\rm
        \odot}\\0.310\times m^{-1.7}\ \ m\ge1\,{\rm M}_{\rm
        \odot}
\end{array}\right.
\end{eqnarray}
\item \citet{chabrier03}:
\begin{eqnarray}
  \xi(m)=\left\{\begin{array}{l}0.497\times m^{-0.2}\ \ 0.1\,{\rm M}_{\rm
        \odot}\le m< 0.3\,{\rm M}_{\rm
        \odot}\\0.241\times m^{-0.8} \ \ 0.3\,{\rm M}_{\rm
        \odot}\le m<1\,{\rm M}_{\rm 
        \odot}\\0.241\times m^{-1.3}\ \ m\ge1\,{\rm M}_{\rm 
        \odot}\end{array}\right.
\end{eqnarray}
\end{itemize}
In the equations above, $\xi(m) = {\rm d}\,{\rm N}/{\rm d} \log m$ describes the number density of stars per logarithmic mass interval. \\

 To identify the collapsed structures in a simulation we use an on the fly parallel Friends-of-Friends (FoF) algorithm. Following \citet{Dolag2009}, we use a linking length of 0.16 times the mean DM particle separation. The star formation rate of a collapsed object is the instantaneous SFRs of gas particles at the current time-step.  The total stellar mass is the sum of the particle stellar masses in the structure. 

\subsection{Feedback models}
\label{subSup}

In our simulations both SN driven galactic winds and AGN feedback are included. In the following we report the parameters used: the interested reader can find an extensive description in \citet{TescariKaW2013}.

For the supernova driven outflows, we use the original \citet{springel2003} energy-driven implementation of galactic winds. The wind carries a fixed fraction $\chi$ of the SN energy ($\epsilon_{\rm SN} = 1.1\times 10^{49}$ erg$/{\rm M}_{\odot}$ for our adopted Chabrier IMF). We assume a wind mass loading factor $\eta=\dot{\rm M}_{\rm w}/\dot{\rm M}_{\rm \star}=2$ and consider the velocity of the wind $v_{\rm w}$ as a free parameter. We explore three different wind velocities:
\begin{itemize}
\item weak winds: $v_{\rm w}=350$ km/s (corresponding to $\chi=0.22$);
\item strong winds: $v_{\rm w}=450$ km/s  ($\chi=0.37$);
\item very strong winds: $v_{\rm w}=550$ km/s ($\chi=0.55$).
\end{itemize}
In one simulation we also include variable momentum-driven winds:
\begin{eqnarray}
  v_{\rm w}= 2\;\sqrt{\frac{G{\rm M}_{\rm halo}}{R_{\rm 200}}}=2\times v_{\rm circ},
\end{eqnarray}
\begin{eqnarray}
  \eta = 2\times\frac{450\,\,{\rm km/s}}{v_{\rm w}},
\end{eqnarray}
where $v_{\rm circ}$ is the circular velocity, M$_{\rm halo}$ is the halo mass and $R_{\rm 200}$ is the radius within which a density 200 times the mean density of the Universe at redshift $z$ is enclosed \citep{barai13}. As discussed in \citet{TescariKaW2013}, wind particles are hydrodynamically decoupled for a given period of time, to ensure they can effectively escape the central region of galaxies and reach the low density intergalactic/circumgalactic medium. Besides the kinetic feedback just described, contributions from both SNIa and SNII to thermal feedback are considered. \\

In our model for AGN feedback, whenever a dark matter halo, identified by the parallel run-time Friends-of-Friends (FoF) algorithm, reaches a mass above a given mass threshold M$_{\rm th}$ for the first time, it is seeded with a central Super-Massive Black Hole (SMBH) of mass M$_{\rm seed}$ (provided it contains a minimum mass fraction in stars $f_{\star}$).  Each SMBH can then grow by accreting local gas and through mergers with other SMBHs. A fraction of the radiated energy associated to the accreted mass is thermally coupled to the surrounding gas. We consider two regimes for AGN feedback, where we vary the minimum FoF mass M$_{\rm
  th}$ and the minimum star mass fraction
$f_{\star}$ for seeding a SMBH, the mass of the seed M$_{\rm seed}$ and the
maximum accretion radius $R_{\rm ac}$. We define:
\begin{itemize}
\item early AGN formation: M$_{\rm th}=2.9\times10^{10}$ M$_{\rm
    \odot}/h$, $f_{\star}=2.0\times10^{-4}$, M$_{\rm seed}=5.8\times
  10^{4}$ M$_{\rm \odot}/h$, $R_{\rm ac}=200$ kpc/$h$;
\item late AGN formation: M$_{\rm th}=5.0\times10^{12}$ M$_{\rm
    \odot}/h$, $f_{\star}=2.0\times10^{-2}$, M$_{\rm seed}=2.0\times
  10^{6}$ M$_{\rm \odot}/h$, $R_{\rm ac}=100$ kpc/$h$.
\end{itemize}
We stress that the radiative efficiency ($\epsilon_{\rm r}$) and the feedback efficiency ($\epsilon_{\rm f}$) are assumed to be the same in the two regimes. However, in the early AGN configuration we allow the presence of a black hole in lower mass halos, and at earlier times.  We extended our simulations to lower redshifts and found that this configuration leads asymptotically to the Magorrian relation \citep{magorrian1998}, but accentuates the effect of AGN in low mass haloes at high $z$.

\subsection{Outline of simulations}
\label{simout}

\begin{figure*} 
\centering
\vspace{0.23cm}
\includegraphics[scale=0.65]{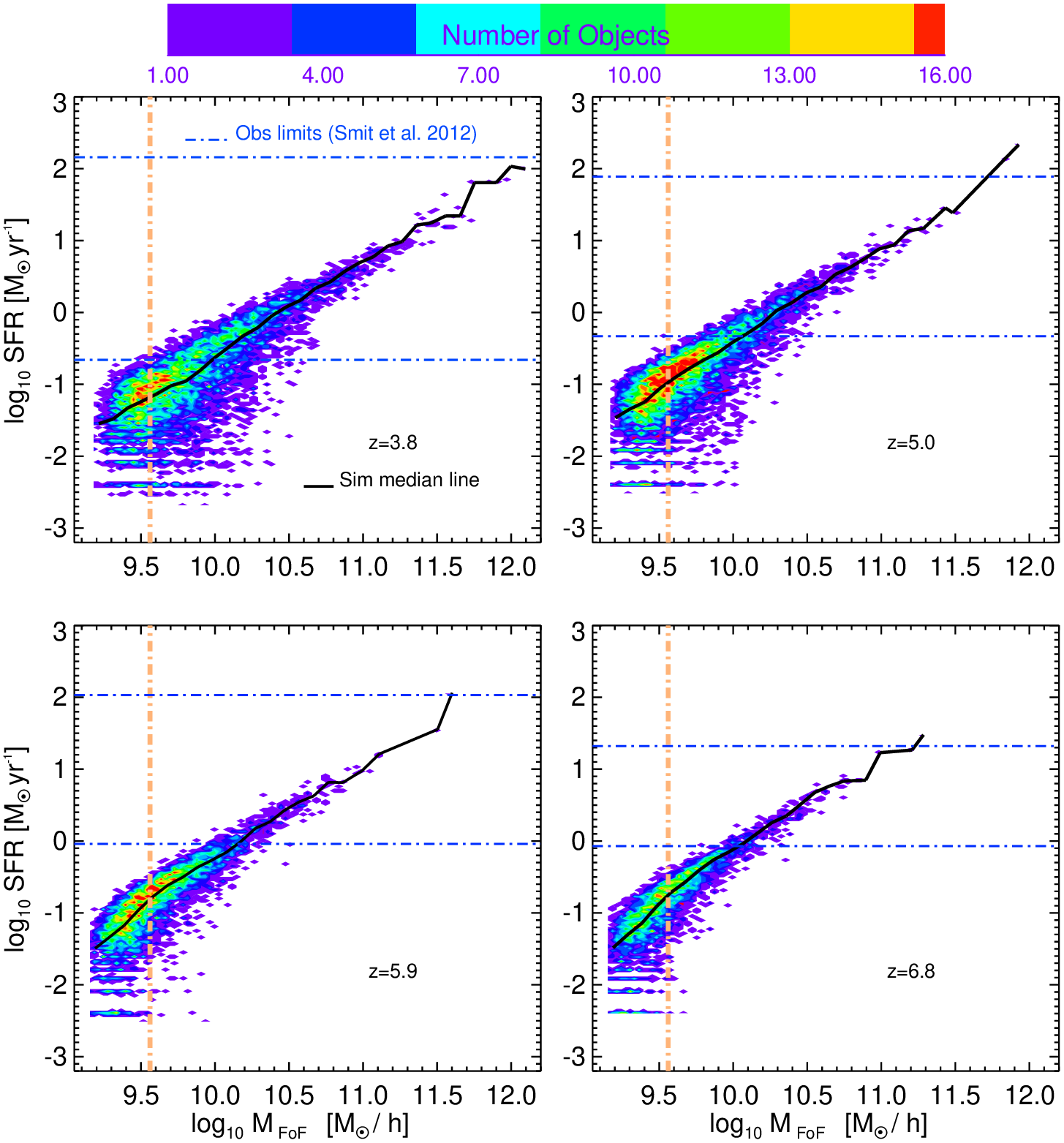} 
\vspace{0.36cm}
\caption{ Density plots of the SFR$-$total halo mass relation for the \textit{Kr24\tu eA\tu sW} run at redshifts $z\sim4-7$. The blue dot-dashed horizontal lines represent the observational limits in the range of SFR of \citet{smit12}. The orange dot-dashed vertical line is our mass confidence limit (see Section \ref{SFR-M}). The black solid line is the median line through the points of the density plots.}
\label{fig:kr24_SFRMass}
\end{figure*}

In Table \ref{tab:sim_runs} we summarise the main parameters of the cosmological simulations performed for this work. Our reference configuration has box size $L = 24$ Mpc/$h$, initial mass of the gas particles M$_{\rm GAS}=7.32\times 10^6$ M$_{\rm \odot}/h$ and a total number of particles (N$_{\rm TOT} =$ N$_{\rm
    GAS}$ $+$ N$_{\rm DM}$) equal to $2\times288^3$. We also ran a simulation with $L = 18$ Mpc/$h$ and one with $L = 12$ Mpc/$h$ to perform box size and resolution tests. All the simulations start at $z=60$ and were stopped at
$z=2$. In the following we outline the characteristics of each run:
\begin{itemize}
\item {\bf{\textit {Kr24\tu eA\tu sW}}}: \citet{kroupa93} initial mass
  function, box size $L=24$ Mpc/$h$, early AGN feedback and strong energy-driven galactic winds of velocity $v_{\rm w}=450$ km/s;
\item {\bf{\textit {Ch24\tu lA\tu wW}}}: \citet{chabrier03} IMF, late
  AGN feedback and weak winds with $v_{\rm w}=350$ km/s;
\item {\bf{\textit {Sa24\tu eA\tu wW}}}: \citet{salpeter55} IMF,
  early AGN feedback and weak winds with $v_{\rm w}=350$ km/s;
\item {\bf{\textit {Ch24\tu eA\tu sW}}}: Chabrier IMF, early AGN
  feedback and strong winds with $v_{\rm w}=450$ km/s;
\item {\bf{\textit {Ch24\tu lA\tu sW}}}: Chabrier IMF, late AGN
  feedback and strong winds with $v_{\rm w}=450$ km/s;
\item {\bf{\textit {Ch24\tu eA\tu vsW}}}: Chabrier IMF, early AGN
  feedback and very strong winds with $v_{\rm w}=550$ km/s;
\item {\bf{\textit {Ch24\tu NF}}}: Chabrier IMF. This simulation was
  run without any winds or AGN feedback, in order to test how large
  the effects of the different feedback prescriptions are;
\item {\bf{\textit {Ch24\tu Zc\tu eA\tu sW}}}: Chabrier IMF, metal
  cooling following \citet{wiersma09}, early AGN feedback and strong winds with $v_{\rm w}=450$
  km/s;
\item {\bf{\textit {Ch24\tu eA\tu MDW}}}: Chabrier IMF, early AGN
  feedback and momentum-driven galactic winds;
\item {\bf{\textit {Ch18\tu lA\tu wW}}}: Chabrier IMF, box size $L=18$
  Mpc/$h$, late AGN feedback and weak winds of velocity $v_{\rm
    w}=350$ km/s. The initial mass of the gas particles is M$_{\rm GAS}=1.30\times10^{6}$ M$_{\rm \odot}/h$, the total number of
  particles is equal to $2\times384^3$;
\item {\bf{\textit {Ch12\tu eA\tu sW}}}: Chabrier IMF, box size $L=12$
  Mpc/$h$, early AGN feedback and strong winds of velocity $v_{\rm
    w}=450$ km/s. The initial mass of the gas particles is M$_{\rm
  GAS}=3.86\times10^{5}$ M$_{\rm \odot}/h$, the total number of
  particles is equal to $2\times384^3$.
\end{itemize}

As already mentioned in the Introduction, we stress that our simulations have been initially calibrated to reproduce the redshift evolution of the observed cosmic star formation rate density. In the first paper of the series \citep{TescariKaW2013} we also tuned the parameters of feedback by comparing with observations of the SFRF at $z\sim 4-7$. In this work we use those constraints to investigate the evolution of the GSMF and SFR$-{\rm M}_{\rm \star}$ relation in the same redshift interval.

\begin{figure*} 
\centering
\includegraphics[scale=0.8]{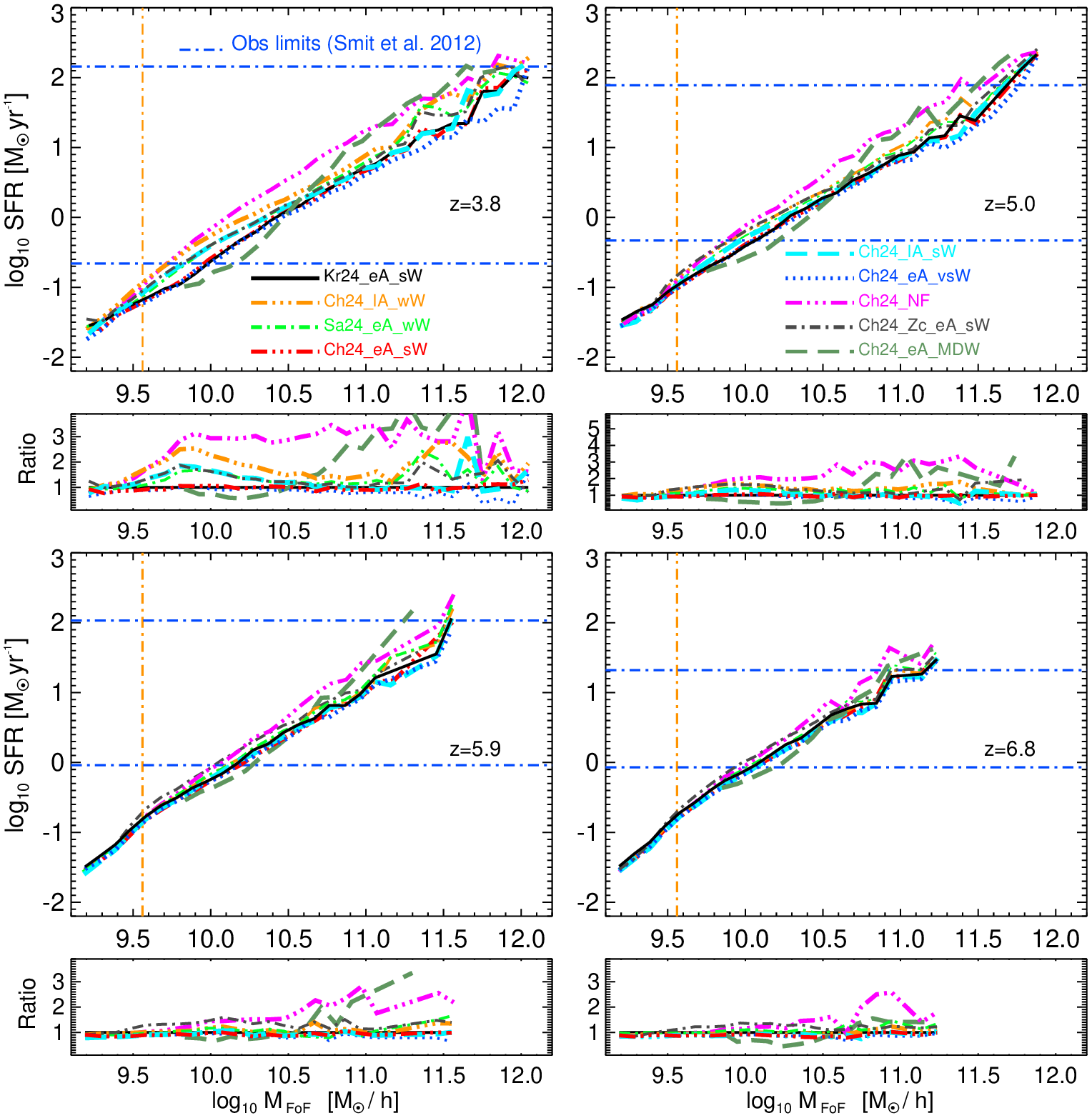}
\vspace{0.33cm}
\caption{SFR$-$total halo mass relation (median lines through density plots) for all the runs of Table \ref{tab:sim_runs} with box size equal to 24 Mpc/$h$. The blue dot-dashed horizontal lines represent the observational limits in the range of SFR of \citet{smit12}. The orange dot-dashed vertical line is our mass confidence limit (see Section \ref{SFR-M}). At each redshift, a panel showing ratios between the different simulations and the \textit{Kr24\tu eA\tu sW} run (black solid line) is included.}
\label{fig:SFR-mass24}
\end{figure*}

\begin{figure*} 
\centering
\includegraphics[scale=0.83]{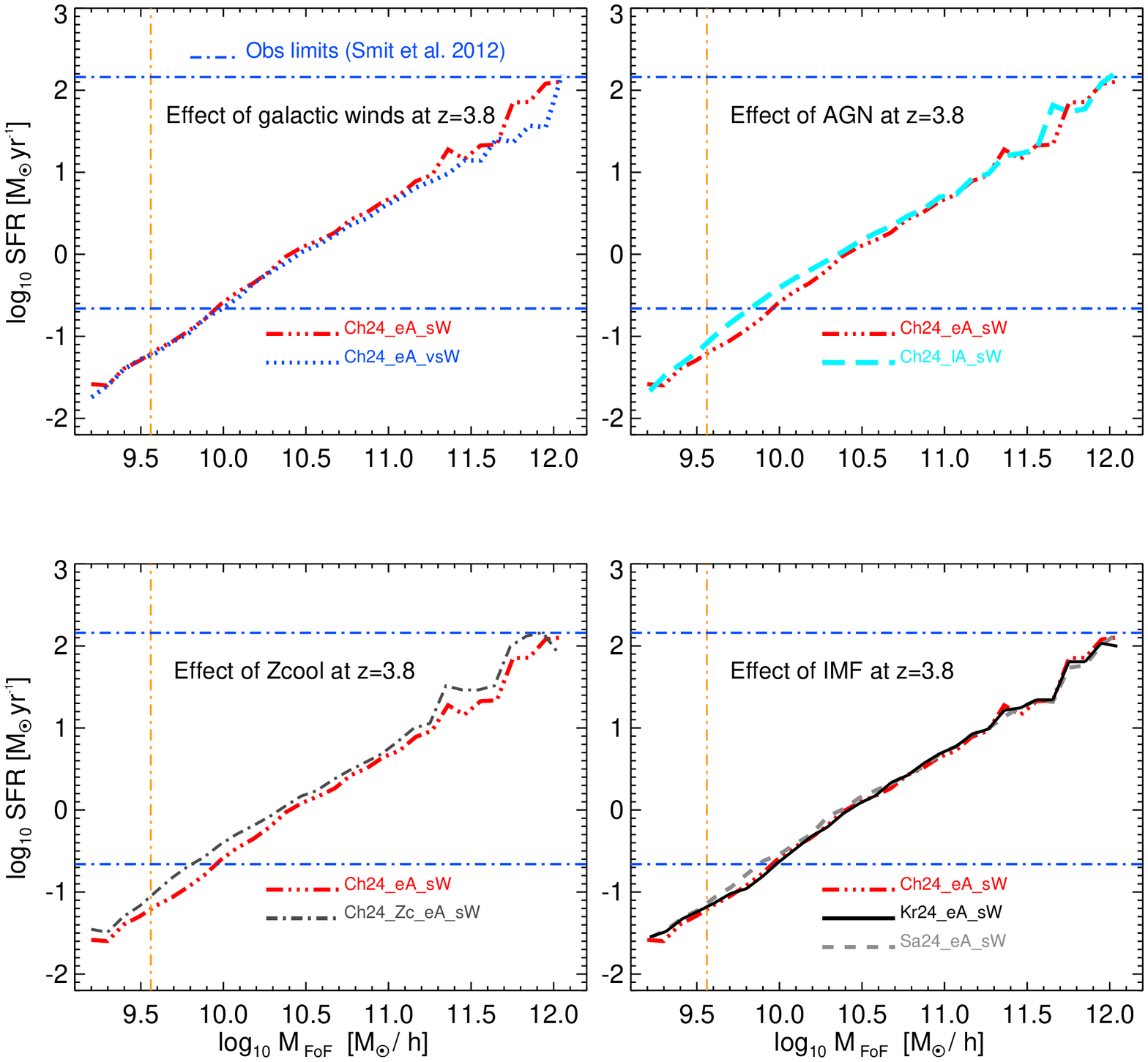}
\caption{SFR$-$total halo mass relation (median lines through  density plots) at $z=3.8$. We evaluate the effects of galactic winds (top left panel), AGN feedback (top right panel), metal cooling (bottom left panel) and choice of IMF (bottom right panel). The blue dot-dashed horizontal lines represent the observational limits in the range of SFR of \citet{smit12}. The orange dot-dashed vertical line is our mass confidence limit (see Section \ref{SFR-M}).}
\label{fig:SFR-mass24clear}
\end{figure*}

We ran all the simulations using the {\textit {raijin}}, {\textit {vayu}} and {\textit{xe}} clusters at the National Computational Infrastructure (NCI) National Facility\footnote{http://nf.nci.org.au} at the Australian National University (ANU). For the post-processing we also used the {\textit {edward}} High Performance Computing (HPC) cluster at the University of Melbourne\footnote{https://edward-web.hpc.unimelb.edu.au/users/}.

\section{Star formation rate in halos}
\label{obdatasampSFRSMall}

\subsection{The star formation rate${\boldsymbol -}$total halo mass relation}
\label{SFR-M}

In this section, we investigate the star formation rate$-$total halo mass relation (SFR$-{\rm M_{FoF}}$) at redshifts $z\sim4-7$. Fig. \ref{fig:kr24_SFRMass} shows the density plot for our fiducial model with a \cite{kroupa93} IMF, early AGN feedback and strong energy-driven galactic winds with constant velocity $v_{\rm w}=450$ km/s. In this figure, the orange dot-dashed vertical lines mark our mass resolution limit. This value is equivalent to the mass of a halo resolved with 100 dark matter particles (M$_{\rm FoF} \sim10^{9.56} $ M$_{\rm \odot}/h$). We showed in Tescari et al. (2014) that the cosmic star formation rate density in structures with larger masses is convergent at redshift $z\le7$, making our results numerically robust in this range. In that paper, we investigated the star formation rate functions of high redshift galaxies and compared these with recent observations from \citet{smit12}. In this paper, we study different galaxy properties, but we mainly compare with the same sample of rest frame UV-selected galaxies from \citet{bouwens2007,bouwens2011}. For this reason we also overplot the observational limits in the range of SFR of \citet[][in Fig. \ref{fig:kr24_SFRMass} blue dot-dashed horizontal lines]{smit12}. In the comparison with observations, we consider only systems inside these observational windows. We stress that halos inside this range are almost always above the mass confidence limit. Only a few are below the mass threshold at redshift $z=3.8$ and are therefore rejected from the subsequent analysis. The black solid lines are the median lines through the points of each density plot. A clear positive correlation between halo mass and SFR is visible, even if the amount of the scatter increases at low mass (especially at lower redshift).

The relation between SFR and total halo mass is not observable, but we explore it to evaluate the effects of different feedback mechanisms and IMFs, and the impact of metal cooling in our simulations. In Fig. \ref{fig:SFR-mass24}, we compare the median lines of the SFR($\rm M_{FoF}$)  density plots for all the runs with box size $L=24$ Mpc/$h$. At each redshift, a panel showing ratios between the different simulations and the \textit{Kr24\tu eA\tu sW} run (black solid line) is included. At redshift $z=6.8$, we see that different configurations are not distinct, except for the no-feedback (\textit{Ch24\tu NF} - magenta triple dot-dashed line) and the Momentum-driven wind (\textit{Ch24\tu eA\tu MDW} - dark green dashed line) cases. By comparing the \textit{Ch24\tu NF} run with all the other simulations, we can see that feedback is already in place at $z\sim7$ and lowers the SFR at any given halo mass.  Momentum-driven winds effectively quench the SFR in low mass halos, but are less efficient than constant winds (with fixed velocity and mass loading factor $\eta=2$) in high mass halos ($\rm M_{FoF}$ $\apprge10^{10.5} $ M$_{\rm \odot}/h$, for the velocity normalisation that we use). This is due to the fact that in this case $\eta$ scales with the inverse of the wind velocity ($\eta\propto v_{\rm w}^{-1}$, see Section \ref{subSup}). The same trends are visible at redshifts $z=5.9$ and $z=5.0$, with the differing behaviour of the \textit{Ch24\tu NF} and \textit{Ch24\tu eA\tu MDW} runs more marked at lower redshift.

Finally, at redshift $z=3.8$, different feedback configurations show different behaviour. The effect of SN driven winds is important for all halos: from the least to the most massive. In the less massive halos ($\rm M_{FoF}$ $\apprle10^{11} $ M$_{\rm \odot}/h$), both strong and weak winds remove gas from the central regions and prevent the formation of new stars. On the other hand, in the most massive halos weaker winds are not able to efficiently expel gas from the high density regions and quench the star formation.

In Fig. \ref{fig:SFR-mass24clear} we highlight the effect of different forms of feedback, metal cooling and IMF on the SFR($\rm M_{FoF}$) relation at $z=3.8$. To explore the impact of galactic winds, in the top left panel we compare the \textit{Ch24\tu eA\tu sW} (red triple dot-dashed line) and the \textit{Ch24\tu eA\tu vsW} (blue dotted line) runs. These runs have exactly the same configuration with the exception of velocity of the winds (450 km/s and 550 km/s, respectively). As argued above, at high masses (M $\apprge10^{11} $ M$_{\rm \odot}/h$), the SFR is lower when the wind velocity is higher (i.e. the blue dotted line is always below the red triple dot-dashed line).

On the other hand, the effect of AGN feedback is particularly visible at low masses (M $\apprle10^{10.5}$ M$_{\rm \odot}/h$). This can be seen by comparing the values of the SFR(M) relation for the \textit{Ch24\tu eA\tu sW} (red triple dot-dashed line) and the \textit{Ch24\tu lA\tu sW} (cyan dashed line) runs (top right panel of Fig. \ref{fig:SFR-mass24clear}). These runs have the same wind strength, but different AGN configurations. In the first case, we adopted the ``early AGN'' scheme, where we reduced the halo mass and the star mass fraction thresholds (M$_{\rm th}$ and $f_{\star}$) to seed a central super-massive black hole of mass M$_{\rm seed}$, with respect to the ``late AGN'' scheme. However, the radiative efficiency ($\epsilon_{\rm r}$) and the feedback efficiency ($\epsilon_{\rm f}$) are the same in the two regimes (see Section \ref{subSup}). \citet{TescariKaW2013} discussed how decreasing the threshold mass for seeding a SMBH increases the effect of AGN feedback on halos with low masses/SFRs, since we impose high black hole$/$halo mass ratios in small galaxies at early times. As a result, at low masses ${\rm SFR}_{\rm\,lA}({\rm M_{FoF}})>{\rm SFR}_{\rm\,eA}({\rm M_{FoF}})$, where the subscripts refer to late and early AGN feedback respectively. At the same time, the effect of AGN feedback on the most massive halos is minimal, since in these objects the central SMBH has not yet reached the self-regulation regime where a lot of energy is released and further star formation prevented.

\begin{figure*}
\centering 
\includegraphics[scale=0.65]{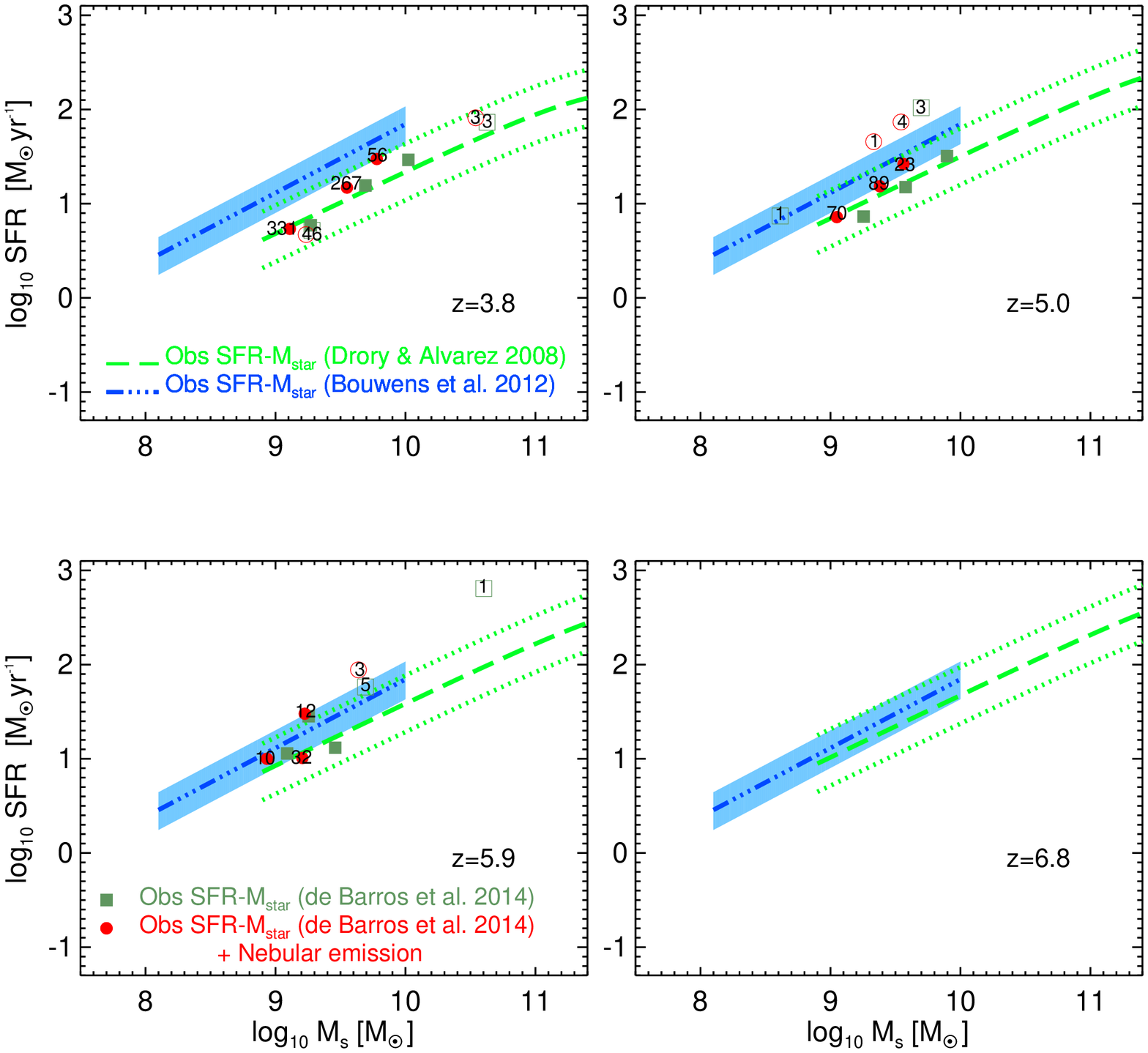}
\caption{The observed galaxy star formation rate$-$stellar mass relation at $z\sim4-7$ from  \citet[][observed frame I-band selected sample - green dashed + dotted lines]{Drory08}, \citet[][drop-outs selection - blue triple dot-dashed lines + light blue shaded regions]{bouwens2012} and \citet[][drop-outs selection - dark green filled squares - No Nebular emission + red filled circles - Nebular emission]{deBarros}.  The open symbols are the median values of \citet{deBarros} that according to the authors suffer from low statistics. The numbers represent the number of galaxies that were used to obtain the median SFR at a given ${\rm M}_{\star}$. The light blue shaded regions represent the statistical error of the SFR$-{\rm M}_{\star}$ relation from \citet{bouwens2012},  estimated by varying the normalization factor in Eq. \ref{eq:SFRparabou}. The green dotted lines represent the 0.3 dex scatter of the SFR$-{\rm M}_{\star}$ relation from \citet{Drory08}.}
\label{fig:obslines}
\end{figure*}

By comparing the \textit{Ch24\tu Zc\tu eA\tu sW} run (dark grey dot-dashed line) with the \textit{Ch24\tu eA\tu sW} run (red triple dot-dashed line) in Fig. \ref{fig:SFR-mass24}, we are able to test how metal cooling affects the SFR$-{\rm M_{FoF}}$ relation. We see that at all redshifts considered, the star formation rate at a given halo mass is always higher in the case of the simulation that includes metal cooling. This is due to the fact that when metals are taken into account in the cooling function, the enriched gas inside halos can cool more efficiently and produce more stars than the same amount of gas of pristine composition. The bottom left panel of Fig. \ref{fig:SFR-mass24clear} shows the comparison between the SFR(${\rm M_{FoF}}$) relations of \textit{Ch24\tu Zc\tu eA\tu sW} and \textit{Ch24\tu eA\tu sW} runs at redshift $z=3.8$.

Since the total integrated amount of gas converted into ``stars'' is the same for different IMFs, for our simulations the choice of initial mass function plays a minor role \citep{TescariKaW2013}. In fact, the run with a Kroupa IMF (\textit{Kr24\tu eA\tu sW} - black solid line) and the simulation with Chabrier IMF (\textit{Ch24\tu eA\tu sW} - red triple dot-dashed line) have almost the same SFR($\rm M_{FoF}$) relation at all halo masses and redshifts. 

In the bottom right panel of Fig. \ref{fig:SFR-mass24clear} we clarify this by showing, in addition to the previous two runs, the SFR(${\rm M_{FoF}}$) relation at $z=3.8$ for a simulation with early AGN, strong winds and \citet{salpeter55} IMF: (\textit{Sa24\tu eA\tu sW} $-$ dashed grey line)\footnote{ This run was used only here to test further the effect of the IMF choice on the simulated SFR(${\rm M_{FoF}}$) relation and it is not part of the set discussed throughout the paper.}. We are comparing models that do not include metal cooling, therefore the source of any difference between these simulations is the number of stars above the SNII mass threshold.

To conclude, we note that at all redshifts considered, when the mass confidence limit is approached (orange dot-dashed vertical lines in Figs. \ref{fig:SFR-mass24} and \ref{fig:SFR-mass24clear}), all the simulations show exactly the same trend because the different feedback configurations start to be numerically poorly resolved. However, we stress that this always happens outside the observational window of \cite{smit12}, which we adopted to perform our analysis. In Appendix A we present the resolution and box size tests using simulations with $L=18$ Mpc/$h$ and $L=12$ Mpc/$h$. These tests show that different configurations are convergent to $\sim0.1$ dex inside the observational limits, with the caveat that simulations with smaller box size are dominated by poor statistics in the high mass end of the SFR$-{\rm M_{FoF}}$ relation.

\subsection{The observed star formation rate${\boldsymbol -}$stellar mass relation}
\label{obdatasampSFRSM}

In this section, we discuss the observed star formation rate$-$stellar mass relation (SFR$-{\rm M}_{\star}$). We use three sets of observations. 

First we consider the results of \citet{bouwens2012}, who derived an approximate relationship between SFR and stellar mass for star-forming galaxies using B-, V-, I- and z-dropouts at $z \sim4$, $5$, $6$ and $7$, respectively. This drop-out technique is a color selection and relies on the Lyman-break in active galaxies. Due to this selection criteria, the sample of \citet{bouwens2012} includes only star forming, mildly obscured galaxies. To obtain the relation between SFR and stellar mass, \citet{bouwens2012} used the observed rest frame UV luminosities, UV-continuum slopes and Mass$/$Luminosity (M$_{\star}/L$) ratios from observations including Ultra-deep ACS and WFC3/IR HUDF+HUDF09 data and wide-area GOODS+ERS+CANDELS data over the CDF-S GOODS field. \citet{bouwens2012} converted UV luminosities into SFRs using the \citet{kennicutt1998} and \citet{madau1998} UV luminosity$-$SFR conversion. A dust correction at different UV luminosities was made using the \citet{meurer1999} (IRX)-$\beta$ relation and the UV-continuum slope, $\beta$, distribution. Then, stellar masses were calculated using the luminosity dependent M$_{\star}/L$ ratios derived by \citet{Gonzalez11}\footnote{\citet{bouwens2012} state that the derived stellar
masses may be up to a factor of 2 higher using mass dependent M$_{\star}/L$ ratios.}.
They found that the SFR$-{\rm M}_{\star}$ relation can be well approximated by the following relation\footnote{The observed SFR$-{\rm M}_{\star}$ relation without dust corrections is:
\begin{eqnarray}
{\rm SFR} = \left(6^{+3}_{-2}\ {\rm M}_{\rm \odot}\,{\rm yr}^{-1}\right)\times\left({\rm M}_{\star}/10^{9}{\rm M}_{\rm \odot}\right)^{0.59 \pm 0.32}\nonumber.
\end{eqnarray}}: 
\begin{eqnarray}
\label{eq:SFRparabou}
{\rm SFR} = \left(13^{+7}_{-5}\ {\rm M}_{\rm \odot}\,{\rm yr}^{-1}\right)\times\left({\rm M}_{\star}/10^{9}{\rm M}_{\rm \odot}\right)^{0.73 \pm 0.32}.
\end{eqnarray}
We plot this relation in the four panels of Fig. \ref{fig:obslines} (blue triple dot-dashed line). The light blue shaded region is the statistical error estimated by varying the normalization factor in Eq. \ref{eq:SFRparabou}. According to \citet{Gonzalez12}, the evolution with redshift of this relationship is very small from $z\sim7$ to $z\sim4$, regardless of the assumed galactic SFH (e.g. a constant or exponentially increasing star formation rate with time). We stress that the SFR$-{\rm M}_{\star}$ relation derived including the effects of dust extinction has a steeper slope than when derived without any dust correction. The intrinsic SFRs calculated by dust correcting the observed SFRs are larger for objects with high SFR/stellar masses. This is due to the fact that dust corrections are more important for big dusty galaxies. Since the intrinsic SFRs are higher at a fixed mass for high star-forming objects, the intrinsic (dust corrected) slope of the SFR$-{\rm M}_{\star}$ relation is steeper. For instance, without taking into account any dust extinction, the results of \citet{Gonzalez11} imply a relationship between SFR and M$_{\star}$ with exponents of 0.59 (i.e. SFR$_{\rm No Dust Corr}\propto {\rm M}_{\star}^{0.59}$). However, when dust correction is included, the SFR(${\rm M}_{\star}$) relation becomes more linear \citep[SFR$_{\rm Dust Corr}\propto {\rm M}_{\star}^{0.73}$,][]{bouwens2012}.

We stress that, according to \citet{bouwens2012} dust corrections increase the normalization of the SFR(${\rm M}_{\star}$) relation by more than a factor of two. This implies that a dwarf galaxy with almost no dust, and stellar mass $\log ({\rm M}{_\star}/{\rm M}_{\odot})\simeq8.5$ at $z\sim4$, should have a SFR $\sim1.9$ times higher if dust corrections are applied\footnote{According to \citet{Sawiki}, $z \sim2$ galaxies with $\log ({\rm M}{_\star}/{\rm M}_{\odot})<9.0$ have a median color excess of $E_{\rm B-V}=0$ (i.e the dust extinction is almost negligible for these objects). \citet{Boquien2012} stress that the use of a canonical starburst ${\rm A_{\rm {FUV}} - \beta}$ relation on normal (non-starburst) star–forming galaxies may produce an overestimate of the SFR by almost an order of magnitude.}. However, \citet{stark2009} pointed out that the inclusion of dust corrections should not lead to an increase in the normalization over time. This could mean that the dust corrections of \citet{bouwens2012} at a fixed stellar mass are overestimated.

The second set of observations we consider comes from \citet{deBarros}. Galaxies at $z \sim3$, 4, 5 and 6 were selected via the presence of the Lyman-break as it is redshifted through the U, B, V, and i bands, respectively. The authors state that their color criteria are very similar to \citet{bouwens2012}. \citet{deBarros} investigated the effect of different choices of the star formation history (constant, decreasing and rising) and nebular emission lines. Stellar masses and SFRs were obtained using Spectral Energy Distribution (SED) fitting taking into account attenuation from intergalactic and interstellar media. For consistency, we consider the constant star formation history  model since \citet{bouwens2012} assumed the same. We plot the results of \citet{deBarros} in Fig. \ref{fig:obslines} (dark green squares). We can see that the inclusion of nebular emission to this model is responsible for a very small shift of the results to smaller masses (red circles of Fig. \ref{fig:obslines}). The results of the authors can be found in table $A_{1}$ of their work. According to \citet{deBarros} there is a very small number of objects in the brightest and faintest bins at each redshift. Therefore, they state that it is considered more appropriate to take into account only the three intermediate bins to examine trends of the physical properties.

\begin{figure}
\centering 
\includegraphics[width=8.5cm]{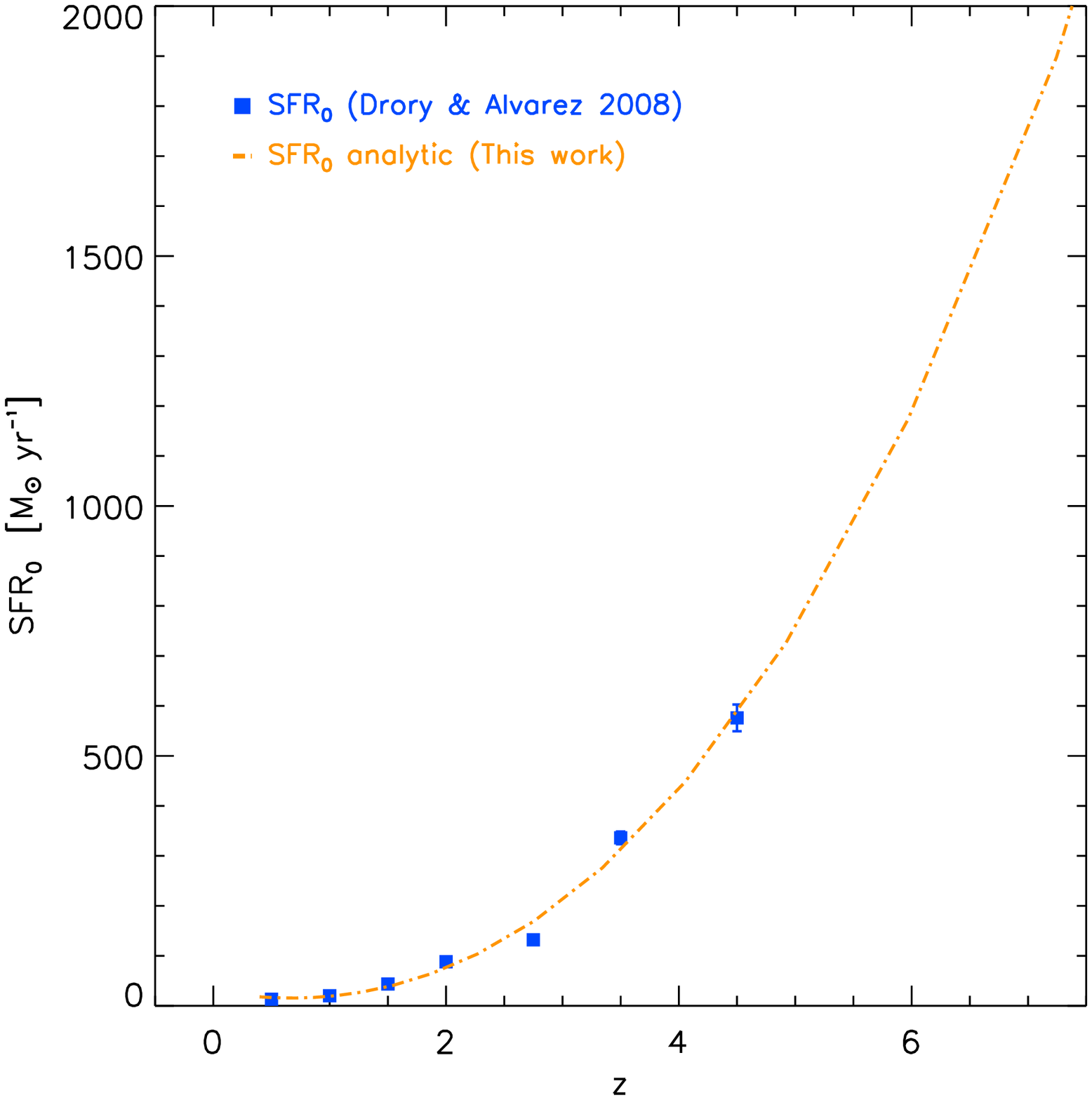}
\caption{Evolution with redshift of the normalization parameter of Eq. (\ref{eq:SFRpar}), SFR$_{\rm 0}$. The blue filled squares with error bars are the observational data from \citet{Drory08}, while the orange dot-dashed line is our fit to the points, extrapolated to $z=7$.}
\label{fig:aparamita}
\end{figure}

The third set of observations that we consider comprises an observed frame I-band selected sample of galaxies. \citet{Drory08} used the data published by \citet[][from the FORS Deep Field and the GOODS-S field]{feulner2005} to find the average star formation rate$-$stellar mass relation at different redshifts. The I-band selected subsample of the FORS Deep Field photometric catalog misses at most $10\%$ of the objects found in ultra-deep K-band observations by \citet{Labbe2003}. As a result \citet{Drory08} were able to include massive and obscured galaxies in their sample.  The stellar masses were computed by fitting models of composite stellar populations with various star formation histories, ages, metallicities, burst fractions and dust contents \citep{Drory2005}.  However, the absence of IRAC photometry can increase the systematic uncertainties in their M$_{\star}$ by a factor of up to 2 \citep{Shapley05}. Then, \citet{Drory08} used dust-corrected UV-continuum emissions to estimate the star formation rates at the same masses.  We stress that the authors used a declining star formation history model to retrieve the SFRs and therefore these can be systematically underestimated by a factor of 5-10 \citep{reddy2012}. However, the SFR$-{\rm M}_{\star}$ relation derived by \citet{Drory08} is in excellent agreement with the recent results of \citet{Salmon2014} who used a constant star formation history model. The dust corrections were determined from stellar population model fits to the multicolor photometry \citep{feulner2005}. The redshift range of the \citet{Drory08} analysis is $0.25<z<5$. The authors parameterised the star formation rate as a power law of stellar mass with an exponential cutoff at high masses:
\begin{eqnarray}
\label{eq:SFRpar}
{\rm SFR} = {\rm SFR}_{0}\,\left(\frac{{\rm M}_{\star}}{{\rm M}_{0}}\right)^{\beta_{\rm s}}\exp\left(-\frac{{\rm M}_{\star}}{{\rm M}_{0}}\right).
\end{eqnarray}
The best fitting parameters are listed in Table 2 of \citet{Drory08}. At redshifts $3<z<4$, the star formation rate is consistent with a single power law with exponent $\beta_{\rm s} \sim 0.6$. Then, at $4<z<5$, the exponent slightly increases to $\beta_{\rm s} \sim 0.65 $. The mass at which the star formation rate starts to deviate from the power law behaviour and becomes an exponential, evolves smoothly from $\log {\rm M}_{0} = 13$ (in units of M$_{\odot}$) at $z\sim5$, to $\log {\rm M}_{0} = 10.5$ (M$_{\odot}$) at $z\sim0.5$. This characteristic mass marks the mass threshold at which the galaxy population changes from actively star-forming at lower stellar masses to quiescent at higher ${\rm M}_{\star}$ \citep{Drory08}. The fact that at $z\ge5$ this characteristic mass is very high, implies that the data are consistent with a single power law, and that there are not many old quiescent massive galaxies. Fig. 4 of \citet{Drory08}, shows the evolution with redshift of the parameter ${\rm M}_{0}$ of Eq. \eqref{eq:SFRpar} and a power-law fit:
\begin{eqnarray}
\label{eq:SFRpara}
{\rm M}_{\rm 0}(z) = 2.66 \times 10^{10} (1+z)^{2.13} \, \, {\rm M}_{\rm \odot}.
\end{eqnarray}
Since we want to test the SFR(${\rm M}_{\star}$) relation at redshifts $4\le z\le7$, we use the fitting parameters of Table 2 in \citet{Drory08} to find the evolution of the normalization parameter ($\rm {SFR}_{0}$) with redshift. Fig. \ref{fig:aparamita} shows this evolution (blue filled squares with error bars) compared with our best fit extrapolated to $z=7$ (orange dot-dashed line):
\begin{eqnarray}
\label{eq:SFRparb}
{\rm SFR}_{\rm 0}(z)=29.5-40.3\,z +28.1\,z^2+1.8\,z^3 \, \, {\rm M}_{\rm \odot}\,{\rm yr^{-1}}.
\end{eqnarray}
As noted, the exponent of the SFR$-{\rm M}_{\star}$ relation is not sensitive to redshift: $\beta_{\rm s}=0.65$. Using Eq. \eqref{eq:SFRpara} and Eq. \eqref{eq:SFRparb} we extrapolate the fitting parameters ${\rm M}_{0}$ and ${\rm SFR}_{0}$, respectively, at redshifts $z\sim$ 4, 5, 6 and 7. We plot the resulting SFR$-{\rm M}_{\star}$ relations in Fig. \ref{fig:obslines} (green dashed + dotted lines). We stress that the observed parameters are determined for redshifts $0.25<z<5$, but we extended this analysis to $z=7$, in order to compare with our simulations. We assume an observed scatter of 0.3 dex regardless of redshift for the SFR$-{\rm M}_{\star}$ relation derived from the observed frame I-band selected sample of \citet{Drory08} \citep[according to][]{Noeske2007,Elbaz07,Daddi2007,Salim2007,Dutton10,Whitaker2012}\footnote{Note that this value may be overestimated. \citet{Whitaker2012} stated that random and systematic errors introduce a 0.18 dex scatter to the SFR$-{\rm M}_{\star}$ relation for star-forming galaxies, from which they estimated the intrinsic scatter to be 0.17 dex.}. 

Fig. \ref{fig:obslines} illustrates the range of SFR$-{\rm M}_{\star}$ relations derived using different sample selections and dust corrections. We can see that the analysis of \citet{Drory08} is able to probe galaxies with higher masses due to their selection criteria. The sample of these authors includes high mass and obscured galaxies\footnote{As noted above, the selection of \citet{Drory08} is effectively a NIR-selection.} that are not found by the drop-outs selection of \citet{bouwens2012} and \citet{deBarros}. The drop-out selection can affect the obtained SFR$-{\rm M}_{\star}$ relation since it takes into account only actively star forming galaxies. \citet{Karim2011} used $1.4$ GHz luminosities and SED fitting to estimate the SFR$-{\rm M}_{\star}$ relation at $z \sim 0.2 - 3.0$ for a mass selected sample. When the authors include only star-forming galaxies in their analysis, the average SFR at a fixed mass is larger, especially at the high mass end of the distribution where quiescent galaxies are present. This effect tends to increase the values of SFR at a fixed mass for the relations obtained by \citet{bouwens2012} and \citet{deBarros}.

 \citet{deBarros} state that their stellar mass$-$luminosity relation is consistent with the one found by \citet{Gonzalez11}. However, the SFR$-{\rm M}_{\star}$ relations of  Fig. \ref{fig:obslines} are quite different. The above studies have similar selection criteria and assume almost the same stellar mass$-$luminosity relation. However, they retrieve star formation rates with different methods. To estimate the SFR at a fixed mass \citet{bouwens2012} used corrections based on UV-continuum slopes \citep{meurer1999} and the \citet{kennicutt1998} relation, while \citet{deBarros} based their results on SED fitting. We see that the star formation rates at a fixed mass are higher in the case of \citet{bouwens2012}. The results of \citet{deBarros} and \citet{Drory08} for the SFR$-{\rm M}_{\star}$ are in good agreement with each other. Both surveys utilised SED fitting for their analysis. \citet{Bauer11} investigated how different methods for the determination of the intrinsic SFR can affect the observed SFR$-{\rm M}_{\star}$ relation. The authors concluded that the best way to determine the amount of dust extinction and the intrinsic SFR is using multiwavelength SED fitting. 

Another difference between the relations shown in Fig. \ref{fig:obslines}, is that \citet{Drory08} calculated the average SFR$-{\rm M}_{\star}$ relation, while \citet{bouwens2012} used the median M$_{\star}/L$ ratios from \citet{Gonzalez11}. Different results are expected using average rather than median M$_{\star}/L$ ratios, since using average ratios would increase the masses. The reason is that in the case of medians, calculations will not take into account the large amounts of stellar mass that reside in the high M$_{\star}$ tail of the mass distribution. 

In conclusion, there is a tension between the SFR$-{\rm M}_{\star}$ relations reported in the literature at $z \sim4-7$. The main reason for this is that observers use different methods and techniques to obtain the intrinsic SFRs and dust corrections. In addition, selection effects can play an important role for the determination of the slope. For instance, Lyman-break selection is biased to detect the most star forming systems at a fixed stellar mass. This effect is more severe for the low mass end  \citep{reddy2012} and high mass end of the distribution \citep{Heinis2014}. This bias increases the normalization of the retrieved SFR$-{\rm M}_{\star}$ relation and possibly makes its slope artificially shallower.

\begin{figure*}
\centering 
\vspace{0.53cm}
\includegraphics[scale=0.65]{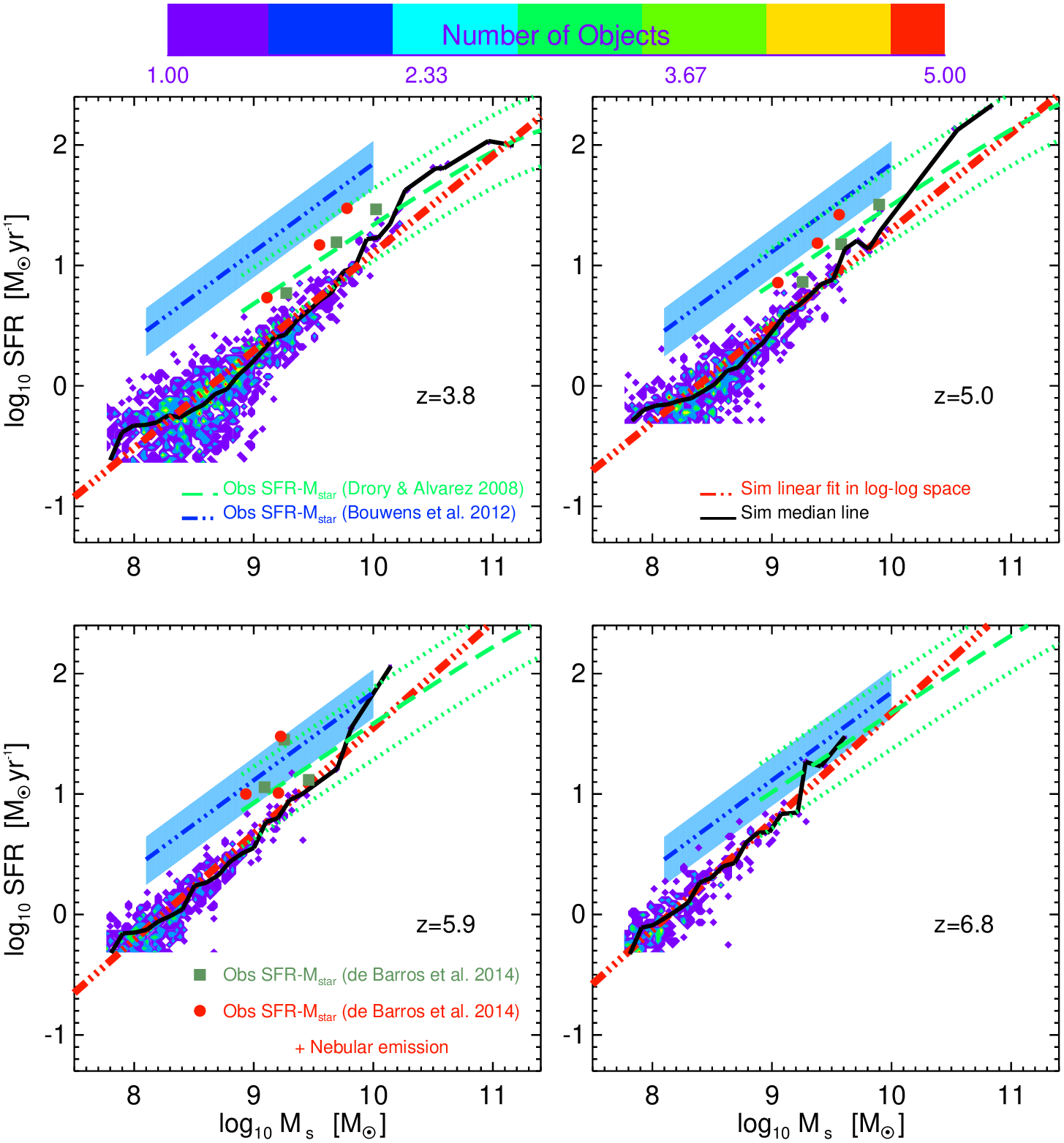}
\vspace{0.53cm}
\caption{ Density plots of the SFR$-$stellar mass relation for our fiducial run \textit{Kr24\tu eA\tu sW} at redshifts $z\sim4-7$. In each panel, the black solid line is the median line through the density plot, while the red triple dot-dashed line is the linear fit (in log-log space,  calculated using a least-square fit) to the points of the  density plot. Overplotted are the observed galaxy SFR(M$_{\star}$) relations from \citet[][observed frame I-band selected sample - green dashed + dotted lines]{Drory08}, \citet[][drop-outs selection - blue triple dot-dashed lines + light blue shaded regions]{bouwens2012} and \citet[][drop-outs selection - dark green filled squares - No Nebular emission + red filled circles - Nebular emission]{deBarros}. The light blue shaded regions represent the statistical error of the SFR$-{\rm M}_{\star}$ relation from \citet{bouwens2012},  estimated by varying the normalization factor in Eq. \ref{eq:SFRparabou}. The green dotted lines represent the 0.3 dex scatter of the SFR$-{\rm M}_{\star}$ relation from \citet{Drory08}.}
\label{fig:kr24_SFRStellarMass}
\end{figure*}

\subsection{Comparison between simulated and observed SFR${\bf -{\rm {\bf M}}_{\star}}$ relation}
\label{CompaSFRSM}

In this section, we compare the observed and simulated SFR$-{\rm M}_{\star}$ relations at high redshift. The evolution of this relation is important since it provides key constraints on the stellar mass assembly histories of galaxies, and also the determination of the galaxy stellar mass function. In Fig. \ref{fig:kr24_SFRStellarMass}, we present a density plot of our fiducial run \textit{Kr24\tu eA\tu sW} at redshifts $z\sim4-7$. In each panel, the red triple dot-dashed line is a  least-square linear fit in log-log space to all the points, and the black solid line their median value in mass bins of 0.1 dex. The halo mass confidence limit of 100 dark matter particles discussed in Section \ref{SFR-M} corresponds to a stellar mass limit of M$_{\star} \sim 10^{7.5}$ M$_{\rm \odot}$. However, in our analysis we take into account objects with M$_{\star}\ge 10^{7.75}$ M$_{\rm \odot}$, which is roughly the mass limit of the current observations \citep{Gonzalez11}. Overplotted are the observed galaxy SFR(M$_{\star}$) relations from \citet[][observed frame I-band selected sample - green dashed + dotted lines]{Drory08}, \citet[][drop-outs selection - blue triple dot-dashed lines + light blue shaded regions]{bouwens2012} and \citet[][drop-outs selection - dark green filled squares - No Nebular emission + red filled circles - Nebular emission]{deBarros}. We find that the simulated relation does not follow either of the three observed relations (and the difference increases at lower redshift where the scatter in the simulated data also increases). In particular, the observed relations are heavily weighted towards high masses while the opposite is true for the simulated relation. We see that the SFR$-{\rm M}_{\star}$ relations obtained from our cosmological hydrodynamic simulations (red triple dot-dashed lines, Fig. \ref{fig:kr24_SFRStellarMass}) have a lower normalization than all the observed relations. This could be due to the fact that observations are unable to detect the faintest and less star forming objects at a fixed stellar mass\footnote{As discussed in the previous section, this is more important for SFR$-{\rm M}_{\star}$ relations that were obtained using Lyman-break selected samples \citep{bouwens2012,Heinis2014}.}. For this reason, the observed  SFR(M$_{\star}$) end up having artificially higher normalization. This bias effect is exacerbated in the low stellar mass tail of the distribution \citep[see e.g. Fig. 26 of][and related discussion]{reddy2012}. Moreover, the intrinsic (bias corrected) observed relation has a slope close to unity, which is in good agreement with cosmological hydrodynamic simulations. In the next paragraphs of this section we discuss in detail the implications of observational biases and different feedback prescriptions on the SFR$-{\rm M}_{\star}$ relation.

\begin{figure*}
\centering 
\includegraphics[scale=0.83]{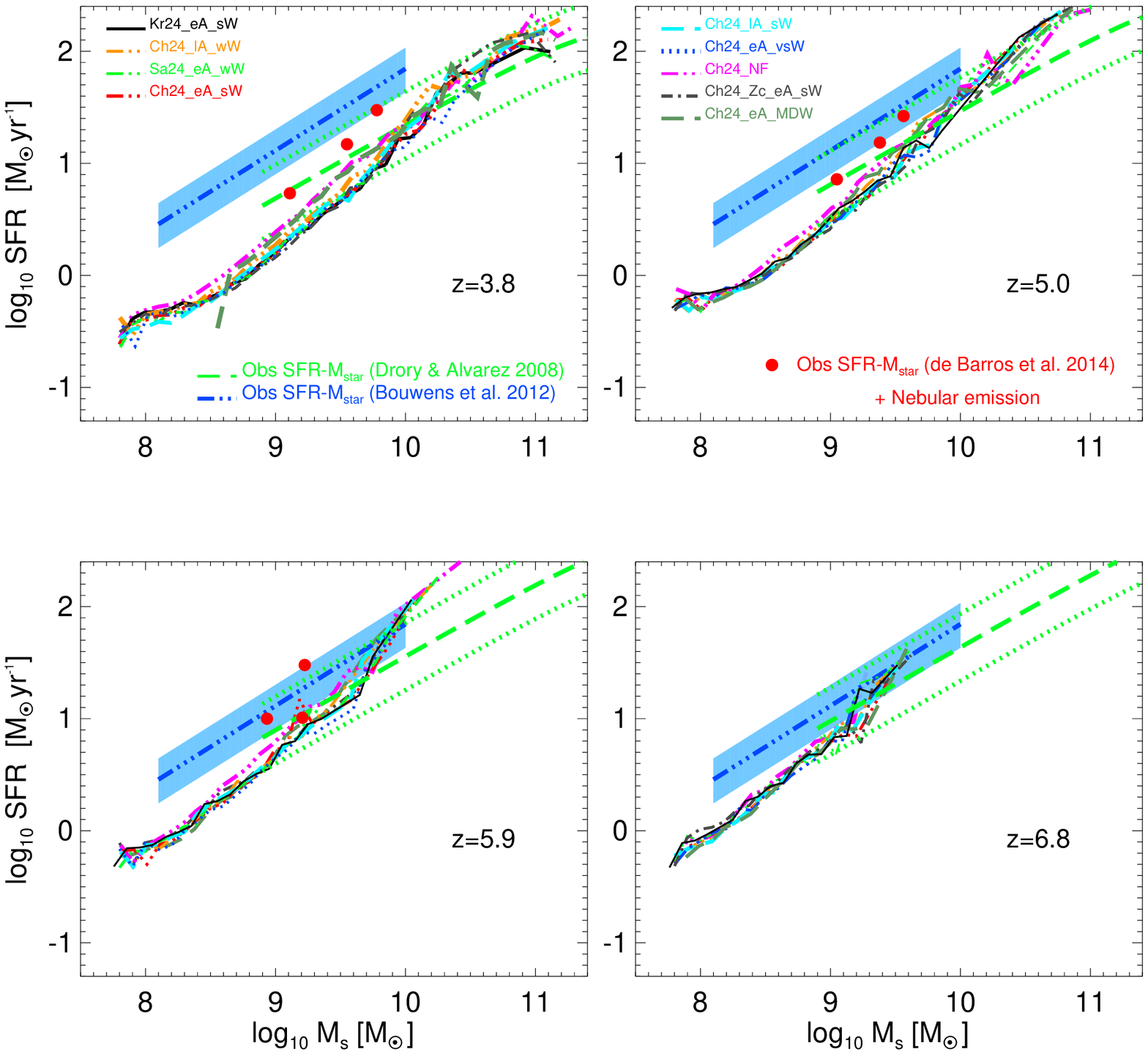}
\caption{Median lines of the star formation rate$-$stellar mass density plots for all the runs of Table \ref{tab:sim_runs} with box size equal to 24 Mpc/$h$. Overplotted are the observed galaxy SFR(M$_{\star}$) relations from \citet[][observed frame I-band selected sample - green dashed + dotted lines]{Drory08}, \citet[][droup-outs selection - blue triple dot-dashed lines + light blue shaded regions]{bouwens2012} and \citet[][droup-outs selection - nebular emission - red filled circles]{deBarros}. Line-styles of different simulations are as in Fig. \ref{fig:SFR-mass24}.}
\label{fig:SFR-StellarMass24}
\end{figure*}

In Fig. \ref{fig:SFR-StellarMass24} we compare the median lines of the SFR$-$stellar mass density plots for all the runs of Table \ref{tab:sim_runs} with box size $L=24$ Mpc/$h$. In Appendix A we will also show the resolution and box size tests using simulations with $L=18$ Mpc/$h$ and $L=12$ Mpc/$h$. These tests demonstrate that the SFR$-{\rm M}_{\star}$ relation converges at all redshifts considered.  At redshift $z=6.8$, the different configurations are not distinct (bottom right panel). The same is true at $z=5.9$ (bottom left panel), except in the case of the \textit{Ch24\tu NF} (magenta triple-dot dashed line), where the absence of feedback leads to slightly higher SFRs at fixed stellar masses. The presence of feedback lowers the SFR at a given M$_{\star}$ at redshifts $z=5$ and $z=3.8$ (top right and left panels, respectively). In our simulations the feedback prescription, choice of IMF and metal cooling all seem to play a small role in the normalization and slope of the simulated SFR$-{\rm M}_{\star}$ relation. Lastly, the momentum-driven winds run is in good agreement with all the constant wind models and the observations.

Star formation is thought to be regulated by the balance between the rate at which the cold gas that creates stars is accreted onto the galaxy, and feedback responsible for quenching the SFR \citep{Fin2008,Dutton10,Bouche10}. Generally, a linear SFR$-{\rm M}_{\star}$ relation is expected from cosmological hydrodynamic simulations \citep{BirnboimDekel03,Dave08,Finlator11b}. This implies a star-forming scenario in which the galaxies build up mass exponentially with time \citep{stark2009,Papov11,Lee11}. It is thought that feedback may affect the slope of the SFR(M${_{\star}}$) relation, but a slope near unity is a generic result of numerical models owing to the dominance of cold mode accretion, which produces rapid, smooth infall \citep{Dave08}. Moreover, according to \citet{schaye10} this is also related to the fact that galaxies form stars in a self-regulated fashion. To interpret the SFR(M$_{\star}$) relation it is therefore important to note that M$_{\star}$ and SFR are both correlated with feedback and cooling. For example, the presence of strong SN feedback lowers the SFR for a galaxy and at the same time lowers its stellar mass with respect to the no feedback run. For this reason, the differences between different feedback configurations are very small for these redshifts. However, a larger discrepancy between SFR$-{\rm M}_{\star}$ relations simulated with different feedback prescriptions is expected for lower redshifts \citep{Haas2013,Haas2013b}.

 The intrinsic SFR(M$_{\star}$) relations in our fiducial simulation \textit{Kr24\tu eA\tu sW} (see the red triple dot-dashed lines in Fig. \ref{fig:kr24_SFRStellarMass}) are: 
\begin{itemize}
\item $z\sim4$:\, SFR $\simeq \left(2.0\,{\rm M_{\rm \odot}\,yr^{-1}}\right)\times\left({\rm M_{\star}/10^{9}M_{\rm \odot}}\right)^{0.81}$, 
\item $z\sim5$:\, SFR $\simeq \left(3.2\,{\rm M_{\rm \odot}\,yr^{-1}}\right)\times\left({\rm M_{\star}/10^{9}M_{\rm \odot}}\right)^{0.80}$,
\item $z\sim6$:\, SFR $\simeq \left(4.7\,{\rm M_{\rm \odot}\,yr^{-1}}\right)\times\left({\rm M_{\star}/10^{9}M_{\rm \odot}}\right)^{0.88}$,
\item $z\sim 7$:\, SFR $\simeq \left(5.9\,{\rm M_{\rm \odot}\,yr^{-1}}\right)\times\left({\rm M_{\star}/10^{9}M_{\rm \odot}}\right)^{0.90}$.
\end{itemize}
We find that the normalization factor evolves from $z=7$ to 4, while the exponent is nearly constant with an averaged value of $\sim 0.85$. The predicted relation from simulations has a slope almost $\sim1$, regardless of the configuration used. \citet{Wilkins13} also find in their simulations that the intrinsic M$_{\star}/L_{\rm UV}-L_{\rm UV}$ relation is almost flat, implying a linear SFR$-{\rm M}_{\star}$ relation. Furthermore, \citet{Finlator11b} found an almost linear relation between star formation and stellar mass. For comparison, \citet{bouwens2012} estimated the dust corrected SFR(M$_{\star}$) relation and found that SFR $\propto {\rm M}_{\star}^{0.73}$ (while the relation without including dust attenuation was found to be SFR $\propto {\rm M}_{\star}^{0.59}$). Thus, dust attenuation plays an important role in the exponent of the SFR$-{\rm M}_{\star}$ relation.

As discussed above, our simulations predict lower values of SFR than observations, especially for objects with small stellar masses. The large difference between observed and predicted SFRs from cosmological hydrodynamic simulations in the literature indicates either that the simulations provide an imcomplete picture of galaxy assembly and cold accretion, or that the observational results are biased. \citet{reddy2012} discussed the implications of various sample biases in the observed SFR$-\rm{M_{\star}}$ relation. The authors determined both SFR and stellar mass for intermediate redshift galaxies ($1.4<z<2.7$) by means of a SED fitting procedure. Using mock samples, SFR and M$_{\star}$ were found to be positively correlated with an intrinsic slope of $\sim 1$. However, a linear least squares fit directly to the observational data, showed a shallower slope of $\sim0.30$. This bias is due to the fact that their observational sample was selected based on UV luminosities (and not stellar masses). As a consequence, galaxies with larger SFRs at a given stellar mass were preferentially selected. According to \citet{reddy2012}  the SFR at $\log ({\rm M}{_\star}/{\rm M}_{\odot})=9.25$ is artificially $\sim  3 - 4$ times larger than the intrinsic value due to this observational selection bias. \citet{bouwens2012} stressed that their analysis is representative of a luminosity-selected sample and that this bias can affect their results. We see that our simulations predict lower values of SFR at a fixed mass than \citet{bouwens2012}, especially for low mass objects. We are in better agreement with the results of \citet{deBarros}. These authors compiled their sample using the same selection criteria as \citet{bouwens2012}, but computed SFRs and stellar masses using the more reliable SED fitting technique.

On the other hand, the I-band selected sample of \citet{Drory08} allows them to include galaxies with larger masses and the selection of stellar mass in obscured objects. However, their magnitude cut results in a sample which is incomplete for objects with low stellar masses/SFRs. From Fig. \ref{fig:SFR-StellarMass24} we see that our simulations are in good agreement with their proposed SFR$({\rm M}_{\star})$ relation. We checked that there is also a good consistency between our simulations and the recent results of \citet{Salmon2014}.  Despite the good consistency between our numerical results and these observations, the simulated relations tend to be steeper.

It is worth mentioning that, although not presented in this paper, the SFR$-{\rm M}_{\star}$ relation proposed by \citet[][rest frame UV-selected sample of galaxies without dust corrections]{stark2009} is in almost perfect agreement with our results. The authors stressed that dust corrections could introduce considerable errors into the observed relation, so they limited their analysis to the relationship between stellar mass and emerging luminosity (i.e the luminosity inferred from the flux that escapes the galaxy without applying any dust corrections). According to \citet{stark2009}, the dust correction law given by \citet{meurer1999} adds a significant random scatter to the observed relation, and possibly cancels any existing trend with luminosity and redshift\footnote{Note that \citet{bouwens2012} used the \citet{meurer1999} relation to dust correct their observed UV luminosities.}. Their results also suggest that dust corrections could have a small impact on the intrinsic SFR at a fixed stellar mass for $z \sim4-7$.

\begin{figure*}
\centering 
\includegraphics[scale=0.83]{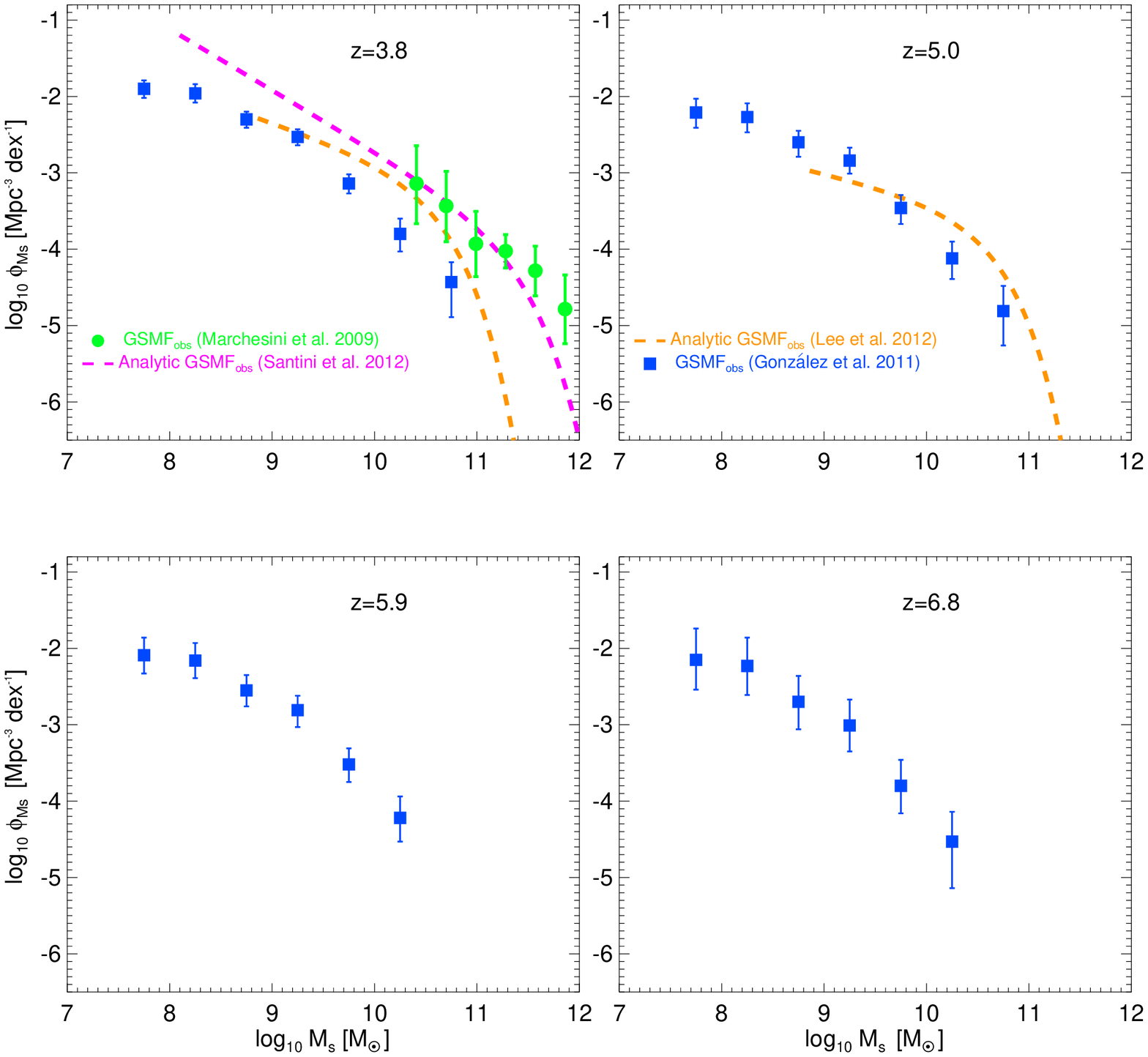}
\caption{Observed galaxy stellar mass functions from \citet[][green filled circles with error bars]{Marchesini09}, \citet[][blue filled squares with error bars]{Gonzalez11}, \citet[][orange dashed lines]{Lee2012} and \citet[][magenta dashed line]{Santini12}.}
\label{fig:aSMFaa}
\end{figure*}

In conclusion, we find that at stellar masses M$_{\star}\apprle10^{9}{\rm M}_{\odot}$ both observations and simulations contain uncertainties, with observations being incomplete and simulations lacking in resolution. For masses M$_{\star}\apprge10^{9}{\rm M}_{\odot}$ our simulations are in agreement with the observations of \citet{Drory08} and \citet{Salmon2014}. In addition, there is a good consistency between our results and the observations of \citet{deBarros}, who used SED fitting to obtain the intrinsic SFRs and stellar masses in their Lyman-break selected sample. Note that these authors were only able to probe a small mass interval ($\log ({\rm M}{_\star}/{\rm M}_{\odot})\sim9.1-9.7$ at $z \sim 4$). Specifically, they do not take into account objects in the low and high mass ends of the distribution, where Lyman-break observations are potentially more affected by selection effects. On the other hand, there is disagreement between our simulations and the observations of \citet{bouwens2012} (especially at $z=3.8$). Our results suggest that their Lyman-break selection may miss a significant number of galaxies, especially at low masses \citep{reddy2012,bouwens2012,Heinis2014}. This conclusion is also supported by the recent work of \citet{wyithe13}. In the next section we show that this incompleteness leads to large uncertainties in the determination of the galaxy stellar mass function.

Finally, our simulations predict a tight SFR$-{\rm M}_{\star}$ relation which is broadly consistent with the results of \citet{Whitaker2012} who found  the scatter of the intrinsic relation to be $\sim$ 0.17.

\section{The galaxy stellar mass function}
\label{Mass function}

\subsection{Observational GSMF}
\label{obdatasampSMF}

In this section we compare our simulations with the observed GSMFs from \citet{Marchesini09}, \citet{Gonzalez11}, \citet{Lee2012} and \citet{Santini12}. \citet{Marchesini09} and \citet{Santini12} used  K-selected galaxy samples and were able to determine the mass function for the most massive galaxies within the redshift range we study. Selecting galaxies from a  K-selected sample includes all galaxies rather than only those that are active and brighty star-forming (as in a  Lyman-break selected sample). However, even the lowest mass galaxies selected by these studies are among the most massive galaxies known at $z>3$ and provide information only about the high end of the GSMF. In addition, these massive galaxies are strongly clustered, which could lead to biased estimations due to the effect of cosmic variance. \citet{Gonzalez11} and \citet{Lee2012} presented GSMFs based on rest frame UV-selected samples, enabling selection of less massive objects. However, the relationships they retrieved are valid only for star-forming galaxies.  In general, the various GSMFs presented in the literature are in good agreement at $z \apprle 3.5$, but differences between studies exist at higher redshift ($z \apprge 4$). In the following we describe each measurement and compare the different observational results. The cosmology assumed is the same in all cases ($H_0=70$ km s$^{-1}$ Mpc$^{-1}$) except for \citet{Lee2012} ($H_0=72$ km s$^{-1}$ Mpc$^{-1}$). Since most of the authors assumed a \citet{salpeter55} IMF, we corrected observed GSMFs to a Salpeter IMF whenever the original choice was different.

\citet{Marchesini09} measured GSMFs of galaxies at redshifts $1.3<z<4.0$. The authors used the combined optical and IR data from 3 different surveys (NIR MUSYC, ultra-deep FIRES and GOODS-CDFS), providing a large area of the sky, so that errors due to cosmic variance should be reduced. In the top left panel of Fig. \ref{fig:aSMFaa}, the green filled circles with error bars show the GSMF of \citet{Marchesini09} at $z\sim4$. We multiplied the stellar masses of \citet{Marchesini09} by 1.6 since the authors used a pseudo-Kroupa IMF instead of a \citet{salpeter55} IMF\footnote{ \citet{Marchesini09} state that they assumed a \citet{salpeter55} IMF for the SED fitting and they divided the derived stellar masses by 1.6 to convert the assumed IMF to a pseudo-Kroupa.}.

 \citet{Santini12} estimated GSMFs in six different redshift intervals between $z\sim0.6$ and $z\sim4.5$ using Early Release Science (ERS) observations taken with the WFC3 in the GOODS-S field. Thanks to deep near-IR observations, they were able to sample the GSMF at masses lower than previous IR studies. However, the authors stated that despite the good agreement with previous work, the limited area of their survey could affect the results at the highest masses (due to cosmic variance). We present the analytic result of \citet{Santini12} in the top left panel of Fig. \ref{fig:aSMFaa} (magenta dashed line). \\

 On the other hand, \citet{Gonzalez11} estimated the GSMF at a given redshift by combining the rest-frame UV Luminosity Function (LF) with the $L_{\rm UV}-{\rm M}_{\star}$ relation measured at $z\sim4$. They used the $z\sim4-7$ UV-LFs from \citet[][Hubble-WFC3/IR camera observations of the Early Release Science field combined with the deep GOODS-S Spitzer/IRAC data]{bouwens2007,bouwens2011}. As a result, the incompleteness-corrected MFs derived by \citet{Gonzalez11} are substantially steeper at low masses than previously found at these redshifts (where incompleteness corrections were not taken into account). The \citet{Gonzalez11} results are shown as the blue filled squares with error bars in Fig. \ref{fig:aSMFaa}.

\citet{Lee2012} investigated how star-forming galaxies typically assemble their mass between redshifts $z \sim4$ and $z \sim5$. We present the fit to the GSMFs for star forming galaxies found in \citet{Lee2012} in the top panels of Fig. \ref{fig:aSMFaa} (orange dashed lines). Since the authors assumed a \citet{chabrier03} IMF, we convert their results to a \citet{salpeter55} IMF by adding 0.16 dex to stellar masses. As noted, the observations from \citet{Gonzalez11} and \citet{Lee2012} (i.e. the UV-selected surveys) do not represent the total SMF of all galaxies at a given redshift, rather they provide information about how much of the cosmic stellar mass density is distributed in actively star-forming and UV-bright galaxies.

The  K-selected  GSMFs from \citet{Santini12} and \citet{Marchesini09} are in agreement  at redshift $z\sim4$. Similarly, \citet{Gonzalez11} is in agreement with \citet{Lee2012} at $z\sim4$, in the mass bin $10^{8.5} \, {\rm M}_{\rm \odot} \apprle {\rm M}_{\star} \apprle 10^{9.5} \, {\rm M}_{\rm \odot}$. At $z\sim5$, the overall normalization of the \citet{Gonzalez11} GSMF is 50\% higher than in \citet{Lee2012}\footnote{At the massive end of the distribution, the data points from \citet{Gonzalez11} are lower by more than a factor of 2 with respect to \citet{Lee2012}. According to \citet{Lee2012}, this is due to the fact that the sample that \citet{Gonzalez11} used covered 75\% less space, and it could therefore be a cosmic variance effect.}. However, where results can  be  directly compared at $z \sim 4$, there is an offset in amplitude between Lyman-break and  K-selected GSMFs.

\subsection{Comparison between simulated and observed galaxy stellar mass functions}
\label{Compsmf}

\begin{figure*}
\centering 
\includegraphics[scale=0.85]{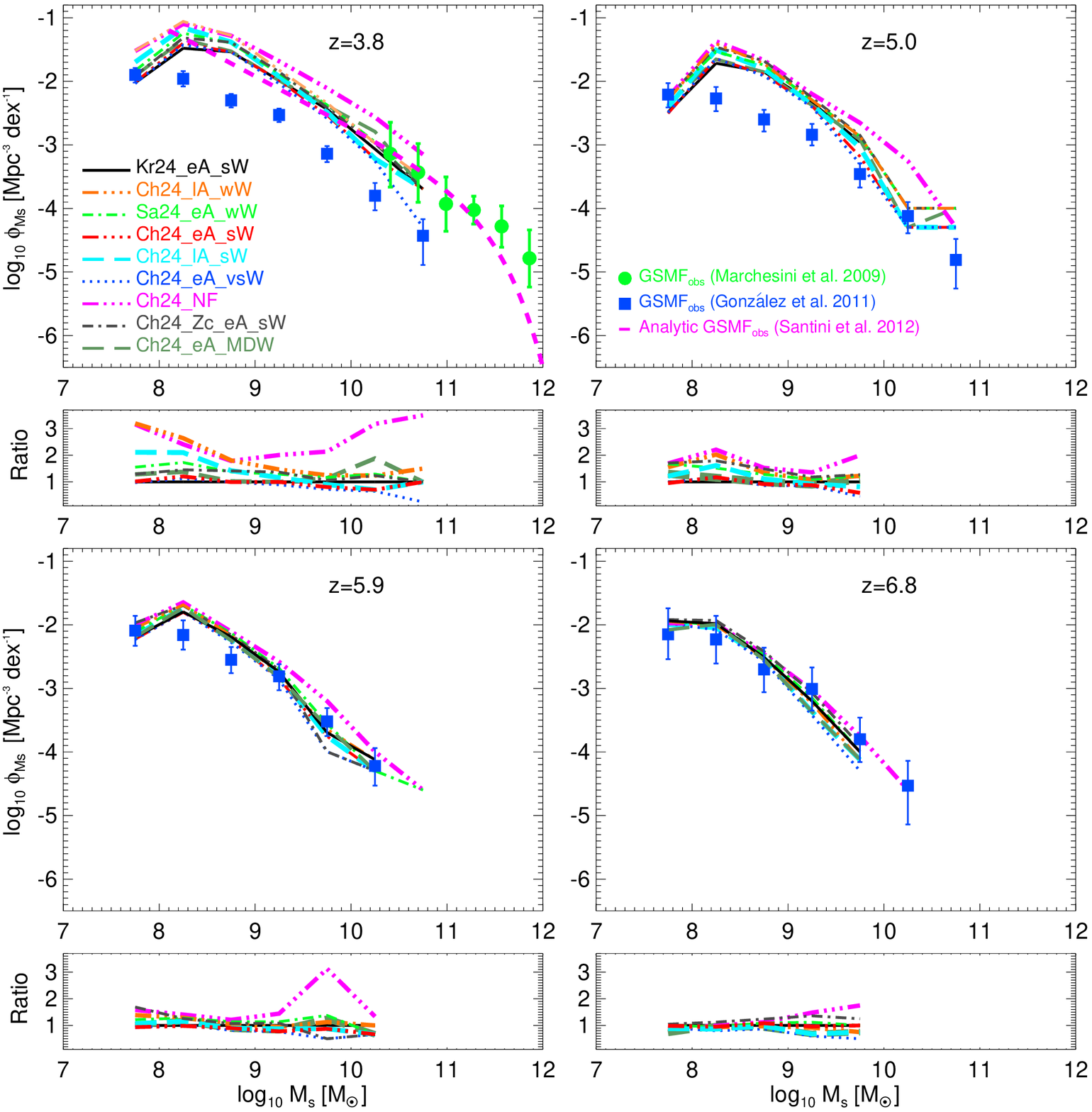}
\caption{Galaxy stellar mass functions for all the runs of Table \ref{tab:sim_runs} with box size equal to 24 Mpc/$h$. Overplotted are the observational results of \citet[][green filled circles with error bars]{Marchesini09}, \citet[][blue filled squares with error bars]{Gonzalez11} and \citet[][magenta dashed line]{Santini12}. At each redshift, a panel showing ratios between the different simulations and the \textit{Kr24\tu eA\tu sW} run (black solid line) is included. Line-styles of different simulations are as in Figs. \ref{fig:SFR-mass24} and \ref{fig:SFR-StellarMass24}.}
\label{fig:aSMFa}
\end{figure*}

We next compare the galaxy stellar mass functions obtained from our simulations with the observed GSMFs discussed in the previous section. Simulations of the GSMF with box size $L=24$ Mpc/$h$ are shown in Fig. \ref{fig:aSMFa}\footnote{Resolution and box size tests using simulations with $L=18$
  Mpc/$h$ and $L=12$ Mpc/$h$ are shown in Appendix A. At each redshift, our simulations show a good convergence down to $\log ({\rm M}{_\star}/{\rm M}_{\odot})=8.25$.}. At each redshift, a panel showing ratios between the different simulations and the \textit{Kr24\tu eA\tu sW} run (black solid line) is included. At $z=6.8$ and $z=5.9$, our results are in agreement with the results from \citet{Gonzalez11}, while at lower redshifts our simulations overpredict the number of star forming selected galaxies of all stellar masses with respect to their observations\footnote{ Our results are instead in excellent agreement with the new observations of \citet{Duncun2014}.}. In addition, at $z<5$ our simulated GSMFs are consistent with the  K-selected GSMFs of \citet{Marchesini09} and \citet{Santini12}. Below we discuss the results at each redshift in more detail.

\begin{figure*}
\centering 
\includegraphics[scale=0.83]{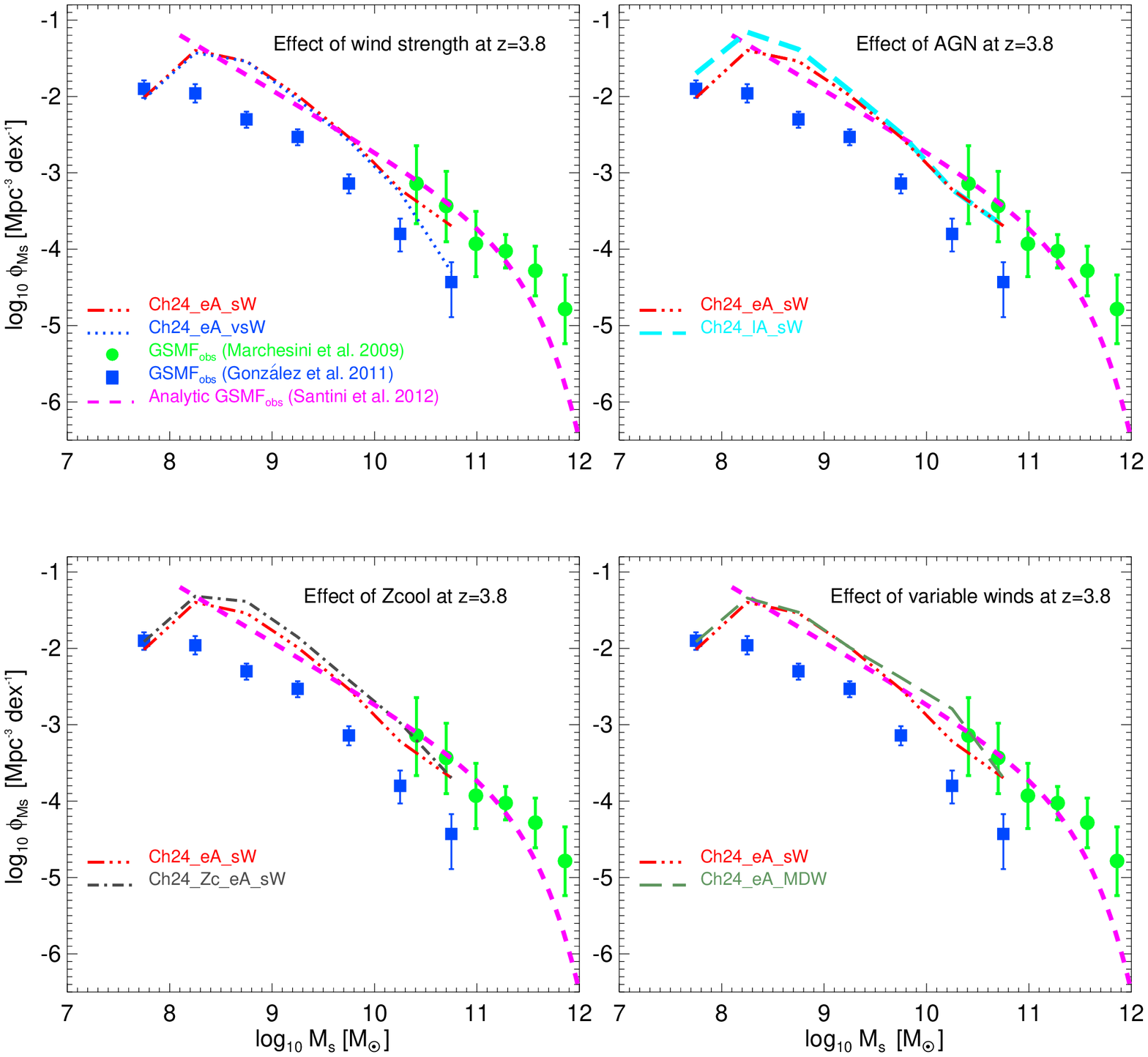}
\caption{Galaxy stellar mass functions at $z=3.8$. We evaluate the effects of galactic wind strength (top left panel), AGN feedback (top right panel), metal cooling (bottom left panel) and momentum-driven winds (bottom right panel) on the GSMF. Overplotted are the observational results of \citet[][green filled circles with error bars]{Marchesini09}, \citet[][blue filled squares with error bars]{Gonzalez11} and \citet[][magenta dashed line]{Santini12}.}
\label{fig:aSMFaclear}
\end{figure*}

At redshift $z=6.8$ (bottom right panel of Fig. \ref{fig:aSMFa}), all simulations show the same trend, regardless the configuration used (strength of feedback, inclusion of metal cooling and choice of IMF). Only the no feedback (\textit{Ch24\tu NF}) and the very strong winds (\textit{Ch24\tu eA\tu vsW}) cases show a small variation with respect to the other runs. This dependence is consistent with the results of \citet{TescariKaW2013} for the SFRF.

At $z=5.9$ (bottom left panel), all simulations are also in good agreement with observations (apart from the no feedback run). In the no feedback case we have an overproduction of systems in the high mass tail of the distribution ($\log({\rm M}_{\star}/{\rm M}_{\odot})\apprge9.3$) with respect to all the other simulations, due to the overcooling of gas.

At $z=5.0$ (top right panel), all  configurations predict more galaxies than observed in almost the whole stellar mass range. Note that the GSMF of the fiducial \textit{Kr24\tu eA\tu sW} run (black solid line) is truncated at $\log({\rm M}_{\star}/{\rm M}_{\odot})\sim9.8$. This is due to low number statistics at higher masses. Therefore, in order to have a fair comparison with all the other simulations, we decided to exclude this region of the \textit{Kr24} GSMF.

The behaviour of different simulations only becomes clear at $z=3.8$ (top left panel). The no feedback case is ruled out by observations of galaxies based on both Lyman-break and  K-band selection. In the following sub-sections and in Fig. \ref{fig:aSMFaclear} we highlight the impact of SN driven galactic winds, AGN feedback, metal cooling and momentum-driven winds on the GSMF at $z=3.8$.

\subsubsection{Effect of SN and AGN feedback}

We start by analysing simulations which have a Chabrier IMF, energy-driven galactic winds and no metal cooling: \textit{Ch24\tu lA\tu wW}, \textit{Ch24\tu lA\tu sW}, \textit{Ch24\tu eA\tu sW} and \textit{Ch24\tu eA\tu vsW}. In the most massive galaxies ($\log({\rm M}_{\star}/{\rm M}_{\odot})\apprge9.8$), the run with late AGN feedback and weak winds (\textit{Ch24\tu lA\tu wW}) produces the highest values of the GSMF. The  runs \textit{Ch24\tu lA\tu sW} and \textit{Ch24\tu eA\tu sW} have strong winds with late and early AGN feedback, respectively. These two simulations do not show any difference at high stellar masses. However, the resulting GSMFs are lower than the \textit{Ch24\tu lA\tu wW} run. Finally, the very strong winds case (\textit{Ch24\tu eA\tu vsW}) has the lowest value of the GSMF in the high mass tail. We highlight the effect of SN feedback in the top left panel of Fig. \ref{fig:aSMFaclear}, by comparing results from the \textit{Ch24\tu eA\tu sW} and \textit{Ch24\tu eA\tu vsW} runs.

The situation is different at low stellar masses ($\log({\rm M}_{\star}/{\rm M}_{\odot})\apprle9.0$). In this range, the \textit{Ch24\tu lA\tu wW} run produces more systems with respect to the other three simulations. However, the \textit{Ch24\tu lA\tu sW} and  \textit{Ch24\tu eA\tu sW} runs are not equal at these low masses. The GSMF of \textit{Ch24\tu eA\tu sW} is lowered by $\sim0.2$ dex. This enhanced effect of AGN feedback, labeled in the low mass end is related to our black hole seeding scheme (see below). The top right panel of Fig. \ref{fig:aSMFaclear} shows the effect of AGN feedback by comparing the \textit{Ch24\tu lA\tu sW} run and the \textit{Ch24\tu eA\tu sW} run.

In \citet{TescariKaW2013} and in Section \ref{SFR-M} we discussed the interplay between SN driven galactic winds and AGN feedback in our simulations. Galactic winds start to be effective before AGN and therefore shape the whole GSMF. In low mass halos the effect of weak and strong winds is the same: both configurations efficiently quench the ongoing star formation by removing gas from the high density central regions. In high mass halos weaker winds become less effective. As a result the  GSMF for {\textit{Ch24\_eA\_vsW}} is lower than the GSMF for {\textit{Ch24\_eA\_sW}} (top left panel of Fig. \ref{fig:aSMFaclear}).

We use two configurations for the AGN feedback, labeled ``early'' and ``late''. In our model, we seed all the halos above a given mass threshold with a central SMBH. In the ``early AGN'' configuration we reduced this threshold and therefore increased the effect of AGN feedback on halos with low mass by construction. For this reason the  GSMF for {\textit{Ch24\_eA\_sW}} is lower than the GSMF for {\textit{Ch24\_lA\_sW}} in the low end of the distribution, $\log({\rm M}_{\star}/{\rm M}_{\odot})\apprle9.0$ (see the top right panel of Fig. \ref{fig:aSMFaclear}). At high masses, SMBHs have not yet reached a regime of self-regulated growth and therefore do not significanlty affect the GSMF. In fact, the GSMF for {\textit{Ch24\_eA\_sW}} is equal to the GSMF for {\textit{Ch24\_lA\_sW}} for $\log({\rm M}_{\star}/{\rm M}_{\odot})\apprge9.5$. We stress that the radiative efficiency ($\epsilon_{\rm r}$) and the feedback efficiency ($\epsilon_{\rm f}$) are each the same in the two AGN configurations.

\subsubsection{Effect of metal cooling}

The simulation with metal cooling (\textit{Ch24\tu Zc\tu eA\tu sW}), shows an increase of the GSMF at all masses with respect to the corresponding simulation without metal cooling (\textit{Ch24\tu eA\tu sW}). This is due to the fact that when metals are included in the cooling function, the gas can cool more efficiently and produce more stars than gas of primordial composition. The effect of metal cooling is visible in the bottom left panel of Fig. \ref{fig:aSMFaclear}.

\subsubsection{Effect of IMF}

 The choice of IMF plays a minor role in the simulated GSMFs. By comparing the \textit{Kr24\tu eA\tu sW} run (Kroupa IMF) with the \textit{Ch24\tu eA\tu sW} run (Chabrier IMF), we see that the only small difference is at redshift $z=3.8$, where the run with Chabrier IMF results in in slightly less galaxies with M$_{\star}\apprge 10^{10}$ M$_{\rm \odot}$ (top left panel of Fig. \ref{fig:aSMFa}).

\subsubsection{Constant- vs. momentum-driven galactic winds}
\label{envsmom}

The physics of the SNe feedback remains an uncertainty in galaxy formation modelling \citep{springel2003,oppe06,tescari09,choina11,tex11,PuchweinSpri12}, and  we have computed results using two different galactic winds schemes. In Fig. \ref{fig:aSMFa} we compare the evolution of the GSMF for the \textit{Ch24\tu eA\tu sW} run (Chabrier IMF, early AGN feedback and energy-driven galactic winds of constant velocity $v_{\rm w}=450$
km/s - red triple dot-dashed line) and the \textit{Ch24\tu eA\tu MDW}
run (Chabrier IMF, early AGN feedback and momentum-driven galactic
winds - dark green dashed line).

At redshift $z=6.8$, 5.9 and 5.0 we find that the two simulations are in excellent agreement. At $z=3.8$ the only notable difference is a slight overproduction of systems with stellar masses $\log({\rm M}_{\star}/{\rm M}_{\odot})\apprge9.5$ for the \textit{Ch24\tu eA\tu MDW} run (bottom right panel of Fig. \ref{fig:aSMFaclear}). This is due to the fact that momentum-driven winds are less efficient than constant winds in the most massive halos. As we showed in \citet{TescariKaW2013}, at $z\apprge4$ the effect of the two different galactic wind implementations on the star formation rate function is significant, while the two schemes result in almost the same GSMF evolution.

\begin{figure*}
\centering 
\includegraphics[scale=0.83]{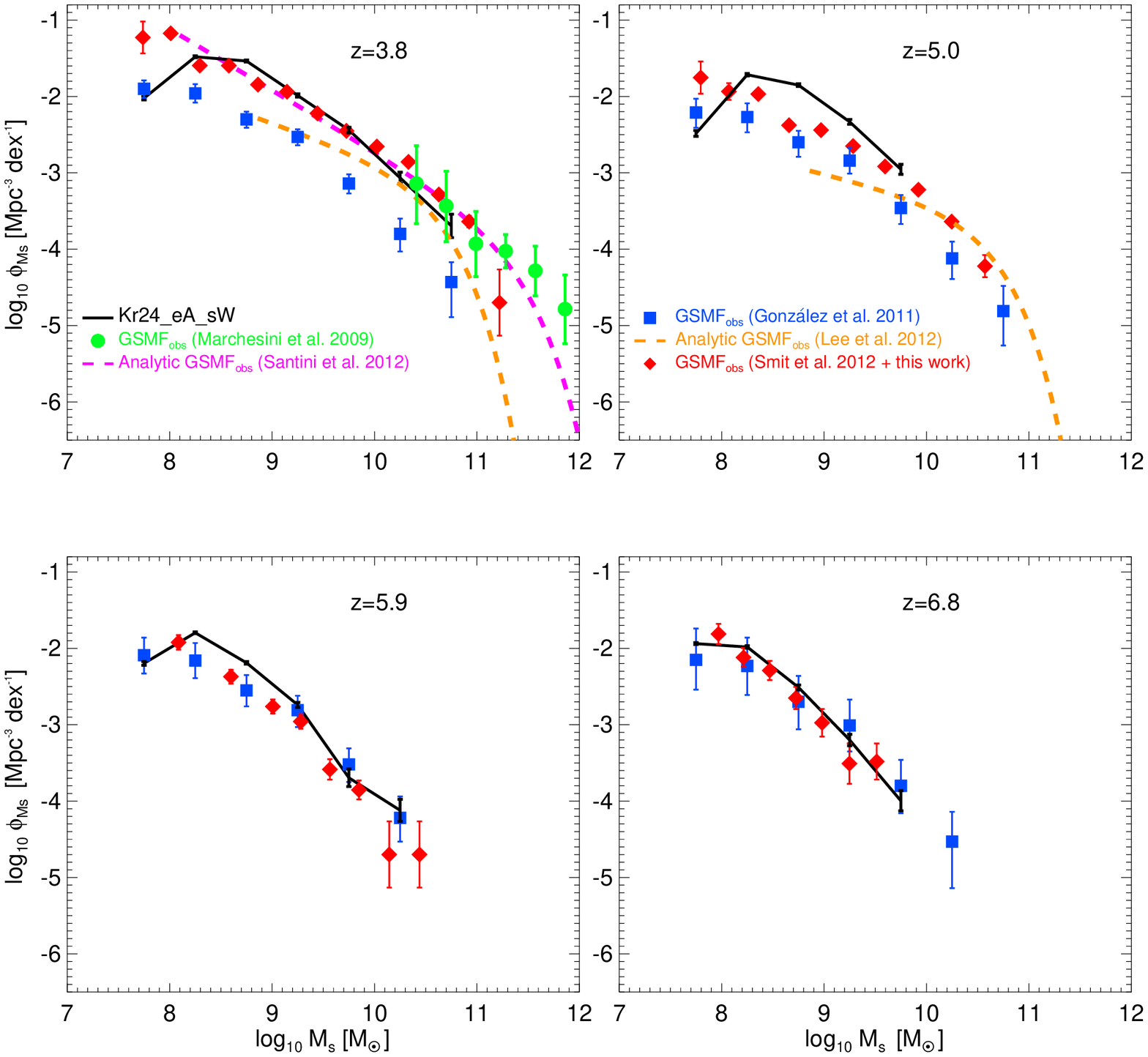}
\caption{Galaxy stellar mass functions at $z\sim4-7$ for our fiducial model with Kroupa IMF, early AGN feedback, constant galactic winds ($v_{\rm  w}=450$ km/s) and box size equal to 24 Mpc/$h$ (black solid lines). The black error bars are the Poissonian uncertainties of the simulated GSMFs. Overplotted are the observational results of \citet[][green filled circles with error bars]{Marchesini09}, \citet[][blue filled squares with error bars]{Gonzalez11}, \citet[][orange dashed lines]{Lee2012} and \citet[][magenta dashed line]{Santini12}. At each redshift, the red diamonds with error bars show the GSMF recovered by combing the stepwise determination of the star formation rate function from \citet{smit12} with our simulated intrinsic SFR$-{\rm M}_{\star}$ relation (see Section \ref{bmdis}).} 
\label{fig:aSMFb}
\end{figure*}

\section{Best model and discussion}
\label{bmdis}

In Fig. \ref{fig:aSMFb} we show the galaxy stellar mass functions at redshifts $z\sim4-7$ for our fiducial model \textit{Kr24\tu eA\tu sW} with Kroupa IMF, early AGN feedback and strong energy-driven galactic winds of constant velocity $v_{\rm w}=450$ km/s (black solid lines). We include the Poissonian uncertainties for the simulated GSMFs (black error bars), in order to provide an estimate of the errors from our finite box size. In this figure we also compare this fiducial model with the observations discussed in Section \ref{obdatasampSMF}.

While at $z>5$ our simulated GSMFs are in good agreement with the results inferred by \citet{Gonzalez11}, at $3.8\le z\le5$ there is disagreement between  their Lyman-break selected GSMF and our numerical results. Our runs predict $\sim3$ times more objects than those found by \citet{Gonzalez11}. However, at redshift $z\sim4$, we are in excellent agreement with the results of \citet{Marchesini09}, \citet{Santini12} (the K-selected samples) and the new data of \citet{Duncun2014}. \citet{Santini12} compared their results with the semi-analytic models of \citet{Menci06}, \citet{Monaco07}, \citet{Wang08} and \citet{Somerville12}, at the same redshift. These semi-analytic  models are also broadly consistent with observations, even though they all underestimate the stellar mass function at the high mass end.

\citet{jaacks12} and \citet{Wilkins13} have previously used cosmological hydrodynamic simulations to model the results of \citet{Gonzalez11}. Both studies also found disagreement with these observations. In particular, \citet{Wilkins13} claimed that the tension between simulated and observed GSMFs arises for two reasons. Firstly, at $z\sim5$ there is a difference in the slope of the UV luminosity function at low $L_{\rm UV}$. According to the authors, this difference is probably due to the fact that feedback in their simulations is not efficient enough for galaxies with M$_{\star}< 10^{9}$ M$_{\rm \odot}/h$. The second reason is that \citet{Gonzalez11} did not take into account the evolution in the assumed $L_{\rm UV}-{\rm M}_{\star}$ relation. As described in Section \ref{obdatasampSMF}, at $4\le z\le7$ the authors used the $L_{\rm UV}-{\rm M}_{\star}$ relation calibrated at redshift $z\sim4$. However, \citet{Wilkins13} found in their simulations that this relation evolves significantly over time and this is partly responsible for the inconsistency between the simulated and the observed GSMF at $z>5$.

In hydrodynamic simulations, the excess of simulated systems at the low mass end of the GSMF is a well known problem \citep{Dave2011}, and is related to the tendency of the SPH simulations to overproduce halos with stellar masses in the range M$_{\star}< 10^{10}$ M$_{\rm \odot}/h$ \citep{Lofaro09}. Once too many small galaxies are produced at high redshift, strong stellar feedback must be invoked to suppress their star formation at lower redshift, so as to recover the correct number density at $z = 0$. The crucial objects that are at the origin of this discrepancy are the small star-forming galaxies at $z > 3$. Moreover, hydrodynamic simulations of the early Universe may yield a SFR$-{\rm M}_{\star}$ relation that is almost unavoidably normalised too low at high $z$, irrespective of their feedback scheme \citep{Haas2013}. This is due to the fact that the first generation of stars can only form once a halo is resolved with enough particles to sample the star forming gas. In general, cosmological hydrodynamic simulations lack the resolution to model the first stars expected to form in these halos and this means that simulated galaxies start to form stars and produce stellar feedback too late. As a result, too many stars form at early times making the galaxies more massive (i.e the SFR$-{\rm M}_{\star}$ relation has lower normalization).

As discussed in Sections \ref{obdatasampSFRSM} and \ref{CompaSFRSM} the star formation rate$-$stellar mass relation recovered from our simulations does not follow the observed one. In general, at a given stellar mass, we predict SFRs that are lower than observed. The combination of these two effects leads to the disagreement between simulated and Lyman-break selected GSMFs in Figs. \ref{fig:aSMFa} and \ref{fig:aSMFb}. As already discussed, observations are heavily weighted towards systems with high masses (i.e. high SFRs). The fact that our simulations are in agreement with ( K-selected) observations in this range, while a different trend is visible at M$_{\star}\apprle 10^{10}$ M$_{\rm \odot}/h$, suggests that we are detecting objects not visible in the current surveys. We therefore argue that the ``true'' normalization of the SFR$-{\rm M}_{\star}$ relation is lower than measured in the case of Lyman-break selected samples of galaxies.

In \citet{TescariKaW2013} we showed that our best fit simulation reproduces the observed star formation rate functions of \citet{smit12} at $z \sim4-7$, determined from the UV-LFs of \citet{bouwens2007}. In light of this result, we tried a simple test to better understand what causes the tension between simulations and observations. At each redshift, we started with the stepwise determination of the SFRF from \citet{smit12} and converted SFRs to stellar masses using our intrinsic simulated SFR$-{\rm M}_{\star}$ relation (see the red triple dot-dashed lines in Fig. \ref{fig:kr24_SFRStellarMass}) to obtain a new estimate of the GSMF. The red diamonds shown in Fig. \ref{fig:aSMFb} represent the results of this test. Note that this approach takes into account the redshift evolution of the SFR$-{\rm M}_{\star}$ relation. We see that the resulting GSMFs are in good agreement with the \textit{Kr24\tu eA\tu sW} run. There is only a small difference at redshift $z=5$, where the simulated SFRF is also slightly different than the observed one \citep{TescariKaW2013}. We find that the determination of the GSMF is extremely sensitive to the choice of the SFR$-{\rm M}_{\star}$ relation. This supports the idea that the reason why our simulations overpredict the observed GSMFs of \citet{Gonzalez11} at $z\le5$ is the inconsistency between the observed and simulated SFR$-{\rm M}_{\star}$ (or $L_{\rm UV}-{\rm M}_{\star}$) relations. At the same time, we also match the observations of \citet{Gonzalez11} at $z\ge6$ better because in this redshift range the SFR(${\rm M}_{\star}$) relations implied by observations are close to the ones from our fiducial simulation (especially the normalization).

\section{Conclusions}
\label{concl}

This paper is the second of a series in which we present the results of the {\textsc{Angus}} ({\textit{AustraliaN {\small{GADGET-3}} early Universe Simulations}}) project. In the first paper of the series \citep{TescariKaW2013} we constrained and compared our hydrodynamic simulations with observations of the cosmic star formation rate density and Star Formation Rate Function (SFRF) for $z \sim4-7$. In this work, we study the relations between star formation rate and both total halo mass and stellar mass (SFR$-{\rm M}_{\rm FoF}$ and SFR$-{\rm M}_{\star}$, repsectively) and investigate the evolution of the galaxy stellar mass function (GSMF) for the same redshift interval. In particular, we have focused on the role of supernova driven galactic winds and AGN feedback. For most of our simulations we used the \citet{springel2003} implementation of SN energy-driven galactic winds. We explored three different wind configurations (weak, strong and very strong winds of constant velocity $v_{\rm w}=350$, 450 and 550 km/s, respectively).  In one simulation we also adopted variable momentum-driven galactic winds. In addition, we explored two regimes for AGN feedback (early and late). The early AGN scheme imposes high black hole$/$halo mass ratios in small galaxies at early times. Finally, we investigated the impact of metal cooling and different IMFs \citep{salpeter55,kroupa93,chabrier03}.

In the following we summarise the main results and conclusions of our analysis:
\begin{itemize}
\item Different feedback prescriptions impact on the SFR$-{\rm M}_{\rm FoF}$ relation. SN driven galactic winds have the most significant effect. The choice of initial mass function plays a minor role in our simulations for this study, while the star formation rate at a given mass is always higher when metal cooling is included.
\item Observational studies report a range of different SFR$-{\rm M}_{\star}$ relations, especially at redshift $z \sim4$. The differences are mostly related to the fact that different groups use different methods to recover intrinsic SFRs. Our results, favour SFRs that are obtained using SED fitting techniques \citep{Drory08,deBarros}. We find that estimations of the SFR using the \citet{kennicutt1998} relation and dust corrections that rely on UV-continuum slopes likely overpredict the SFR at a fixed mass.
\item Our simulated SFR$-{\rm M}_{\star}$ relations are in good agreement with the SED based observations of \citet{Drory08} and \citet{deBarros} for stellar masses M$_{\star}\apprge10^{9}{\rm M}_{\odot}$. However, our simulations, that are calibrated against the observed cosmic star formation rate density and SFRF, do not agree with the Lyman-break selected sample of \citet{bouwens2012}. Our simulations predict a population of faint galaxies not seen by current observations.
\item We reproduce the Lyman-break selected galaxy stellar mass functions of \citet{Gonzalez11} at $z=6.8$ and 5.9. At lower redshift, we are in agreement with the GSMFs determined from K-selected observations in the high mass end of the distribution, but overproduce the number of galaxies with respect to \citet{Gonzalez11}, especially at the low mass end.  Feedback effects are important in reproducing the observed GSMFs. Models without feedback do not describe the observations.
\item At the highest redshifts considered in this work, energy- and momentum-driven galactic winds predict the same SFR$-{\rm M}_{\star}$ relation and the same galaxy stellar mass function evolution.
\item Our simulated GSMF is consistent with the results of \citet{Marchesini09}, \citet{Santini12} and \citet{Duncun2014}. The GSMF of \citet{Gonzalez11} for $z \sim4-7$ is estimated by converting the luminosity function into a stellar mass function, and is therefore highly dependent on the assumed ${\rm M}_{\star}-L_{\rm UV}$ relationship. This relation is uncertain and possibly biased by a range of factors. The observed relation is also heavily weighted towards systems with high star formation rates. We argue that this is the main reason for the difference between simulations and the observed Lyman-break selected GSMF of \citet{Gonzalez11} at $z \sim4$.
\end{itemize}

In conclusion, we argue that the normalization of the observed SFR$-{\rm M}_{\star}$ relation is overestimated by current measures. The range of results from current surveys arise because of the different and uncertain procedures used for obtaining the intrinsic SFRs, and  because current surveys are unable to detect the low SFR/mass objects. Future deep surveys should find a large population of faint galaxies with low stellar masses. This will result in a steeper SFR$-{\rm M}_{\star}$ relation with smaller normalization.

\section*{Acknowledgments}

We would like to thank  the anonymous referee, Paul Angel, Simon Mutch, Kristian Finlator, Martin Stringer, Kyoung-soo Lee and Mark Sargent for insightful discussions and suggestions. The authors would also like to thank Volker Springel for making available to us the non-public version of the {\small{GADGET-3}} code. ET is thankful for the hospitality provided by the University of Trieste and the Trieste Astronomical Observatory, where part of this work was completed. This research was conducted by the Australian Research Council Centre of Excellence for All-sky Astrophysics (CAASTRO), through project number CE110001020. This work was supported by the NCI National Facility at the ANU.

\begin{figure*}
\centering 
\includegraphics[width=13.3cm, height=10.35cm]{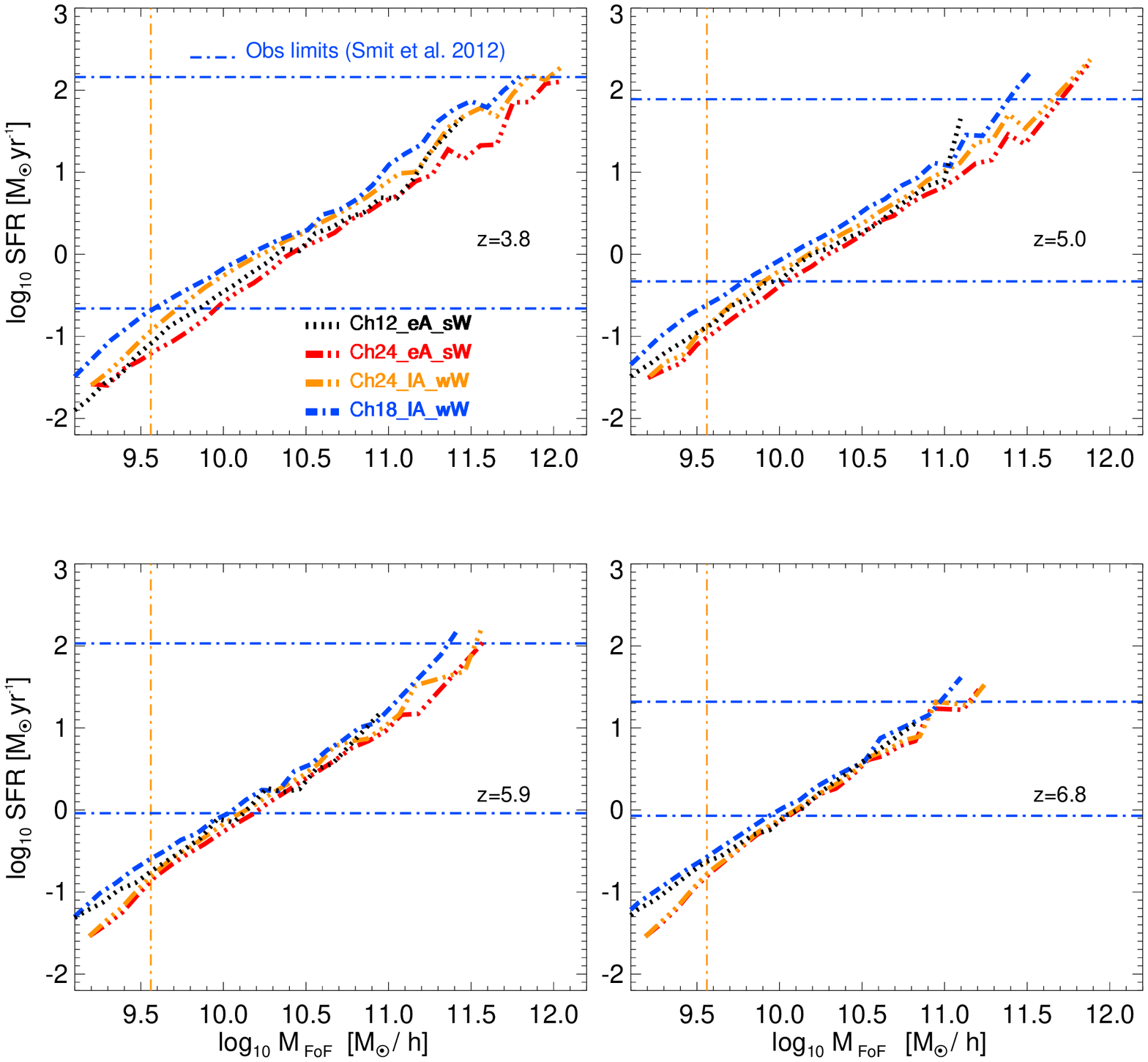}\\ 
\vspace{0.25cm}
\includegraphics[width=13.3cm, height=10.35cm]{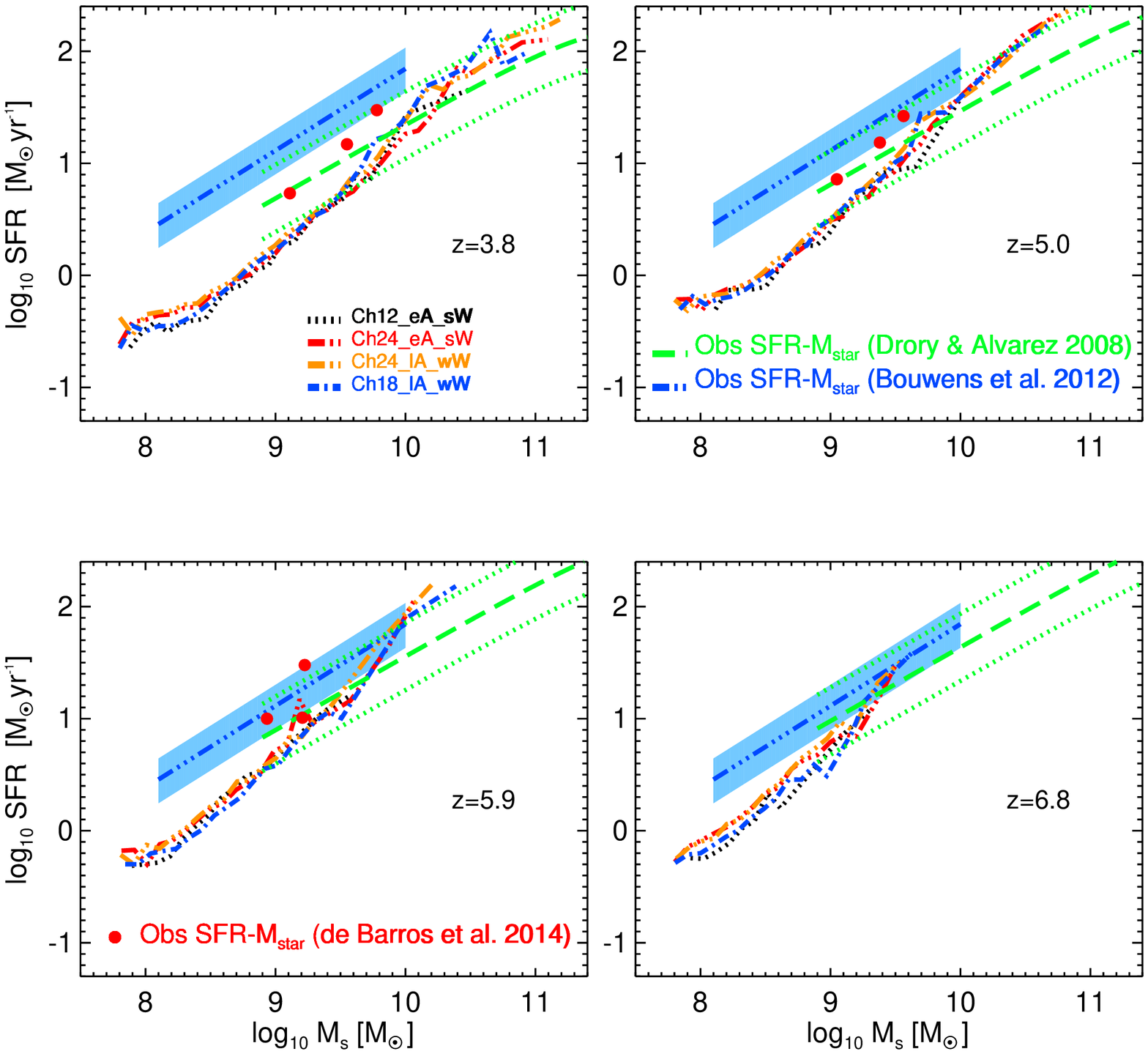}
\caption{Box size and resolution tests at redshift $z\sim4-7$. \textit{Top 4 panels}: SFR$-{\rm M}_{\rm FoF}$ relation. \textit{Bottom 4 panels}: SFR$-{\rm M}_{\star}$ relation. In all the panels we compare $a)$ the Chabrier IMF, Early AGN, Strong Winds configuration for two different box sizes: $L=24$ Mpc/$h$ (\textit{Ch24\tu eA\tu sW} - red triple dot-dashed line) and $L=12$ Mpc/$h$ (\textit{Ch12\tu eA\tu sW} - black dotted line) and $b)$ the Chabrier IMF, Late AGN, Weak Winds configuration for $L=24$ Mpc/$h$ (\textit{Ch24\tu lA\tu wW} - orange triple dot-dashed line) and $L=18$ Mpc/$h$ (\textit{Ch18\tu lA\tu wW} - blue dot-dashed line). In the \textit{top 4 panels} the blue dot-dashed horizontal lines represent the observational limits in the range of SFR of \citet{smit12} and the orange dot-dashed vertical line is our mass confidence limit (see Section \ref{SFR-M}). In the \textit{bottom 4 panels} we overplot the observed galaxy SFR(M$_{\star}$) relations from \citet[][observed frame I-band selected sample - green dashed + dotted lines]{Drory08}, \citet[][droup-outs selection - blue triple dot-dashed lines + light blue shaded regions]{bouwens2012} and \citet[][droup-outs selection - nebular emission - red filled circles]{deBarros}.}
\label{fig:SFR-mass12}
\end{figure*}

\begin{figure*}
\centering
\includegraphics[scale=0.83]{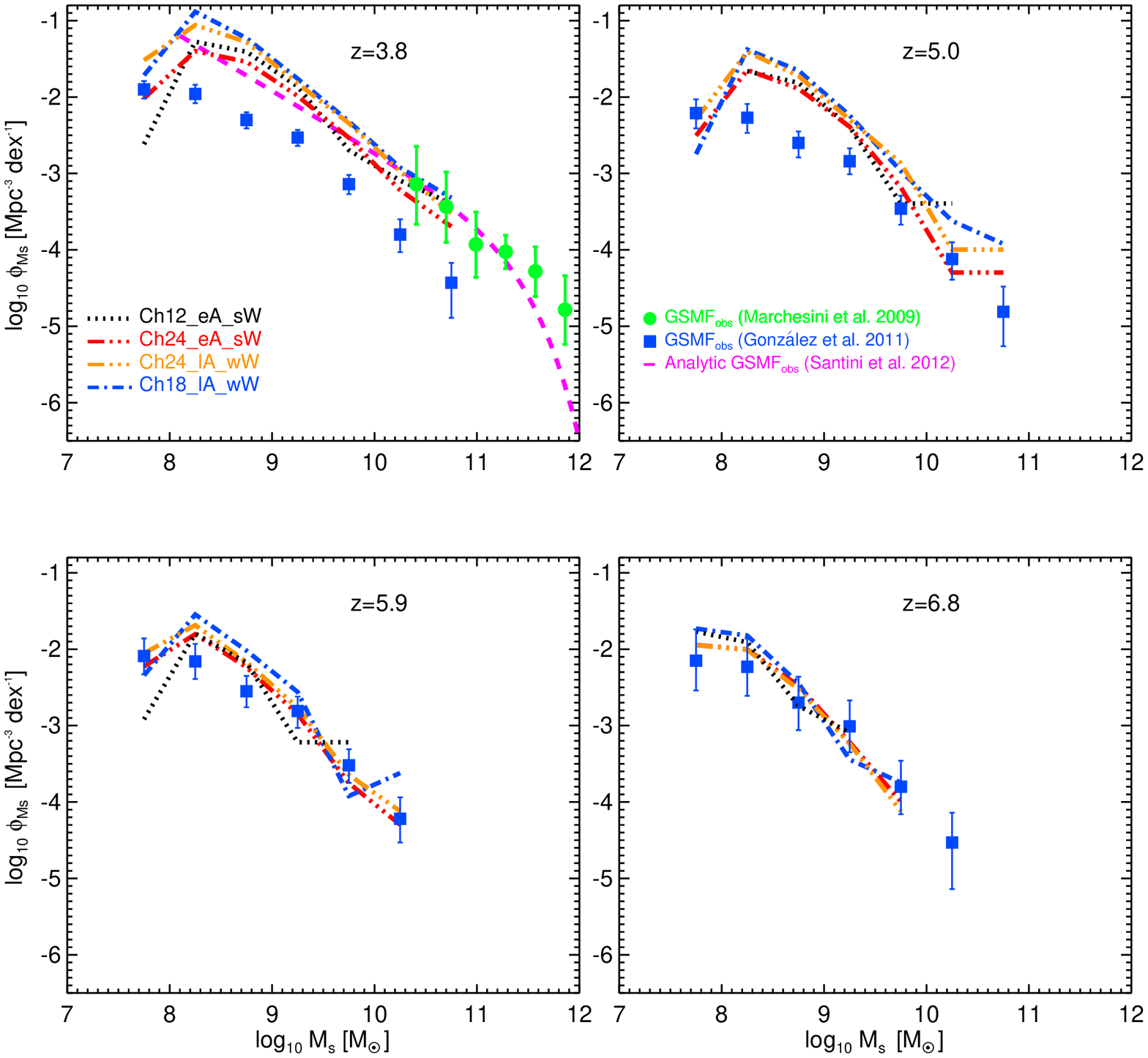}
\caption{Box size and resolution tests: galaxy stellar mass functions at redshift $z\sim4-7$. We compare the same simulations of Fig. \ref{fig:SFR-mass12} (line-styles of different simulations are also the same). Overplotted are the observational results of \citet[][green filled circles with error bars]{Marchesini09}, \citet[][blue filled squares with error bars]{Gonzalez11} and \citet[][magenta dashed line]{Santini12}.}
\label{fig:SMFres}
\end{figure*}

\bibliographystyle{mn2e}	
\bibliography{Katsianis_mnrasRev2}

\newcommand{\noopsort}[1]{}
\begin{thebibliography}{}

\bibitem[\protect\citeauthoryear{{Balogh}, {Christlein}, {Zabludoff} \&
  {Zaritsky}}{{Balogh} et~al.}{2001}]{Balogh01}
{Balogh} M.~L.,  {Christlein} D.,  {Zabludoff} A.~I.,    {Zaritsky} D.,  2001,
  \apj, 557, 117

\bibitem[\protect\citeauthoryear{{Barai}, {Viel}, {Borgani}, {Tescari},
  {Tornatore}, {Dolag}, {Killedar}, {Monaco}, {D'Odorico} \&
  {Cristiani}}{{Barai} et~al.}{2013}]{barai13}
{Barai} P.,  {Viel} M.,  {Borgani} S.,  {Tescari} E.,  {Tornatore} L.,  {Dolag}
  K.,  {Killedar} M.,  {Monaco} P.,  {D'Odorico} V.,    {Cristiani} S.,  2013,
  \mnras, 430, 3213

\bibitem[\protect\citeauthoryear{{Bauer}, {Conselice},
  {P{\'e}rez-Gonz{\'a}lez}, {Gr{\"u}tzbauch}, {Bluck}, {Buitrago} \&
  {Mortlock}}{{Bauer} et~al.}{2011}]{Bauer11}
{Bauer} A.~E.,  {Conselice} C.~J.,  {P{\'e}rez-Gonz{\'a}lez} P.~G.,
  {Gr{\"u}tzbauch} R.,  {Bluck} A.~F.~L.,  {Buitrago} F.,    {Mortlock} A.,
  2011, \mnras, 417, 289

\bibitem[\protect\citeauthoryear{{Birnboim} \& {Dekel}}{{Birnboim} \&
  {Dekel}}{2003}]{BirnboimDekel03}
{Birnboim} Y.,  {Dekel} A.,  2003, \mnras, 345, 349

\bibitem[\protect\citeauthoryear{{Boquien}, {Buat}, {Boselli}, {Baes}
  et~al.,}{{Boquien} et~al.}{2012}]{Boquien2012}
{Boquien} M.,  {Buat} V.,  {Boselli} A.,  {Baes} M.,    et~al., 2012, \aap,
  539, A145

\bibitem[\protect\citeauthoryear{{Bouch{\'e}}, {Dekel}, {Genzel}, {Genel},
  {Cresci}, {F{\"o}rster Schreiber}, {Shapiro}, {Davies} \&
  {Tacconi}}{{Bouch{\'e}} et~al.}{2010}]{Bouche10}
{Bouch{\'e}} N.,  {Dekel} A.,  {Genzel} R.,  {Genel} S.,  {Cresci} G.,
  {F{\"o}rster Schreiber} N.~M.,  {Shapiro} K.~L.,  {Davies} R.~I.,
  {Tacconi} L.,  2010, \apj, 718, 1001

\bibitem[\protect\citeauthoryear{{Bouwens}, {Illingworth}, {Franx} \&
  {Ford}}{{Bouwens} et~al.}{2007}]{bouwens2007}
{Bouwens} R.~J.,  {Illingworth} G.~D.,  {Franx} M.,    {Ford} H.,  2007, \apj,
  670, 928

\bibitem[\protect\citeauthoryear{{Bouwens}, {Illingworth}, {Oesch}, {Franx},
  {Labb{\'e}}, {Trenti}, {van Dokkum}, {Carollo}, {Gonz{\'a}lez}, {Smit} \&
  {Magee}}{{Bouwens} et~al.}{2012}]{bouwens2012}
{Bouwens} R.~J.,  {Illingworth} G.~D.,  {Oesch} P.~A.,  {Franx} M.,
  {Labb{\'e}} I.,  {Trenti} M.,  {van Dokkum} P.,  {Carollo} C.~M.,
  {Gonz{\'a}lez} V.,  {Smit} R.,    {Magee} D.,  2012, \apj, 754, 83

\bibitem[\protect\citeauthoryear{{Bouwens}, {Illingworth}, {Oesch},
  {Labb{\'e}}, {Trenti}, {van Dokkum}, {Franx}, {Stiavelli}, {Carollo}, {Magee}
  \& {Gonzalez}}{{Bouwens} et~al.}{2011}]{bouwens2011}
{Bouwens} R.~J.,  {Illingworth} G.~D.,  {Oesch} P.~A.,  {Labb{\'e}} I.,
  {Trenti} M.,  {van Dokkum} P.,  {Franx} M.,  {Stiavelli} M.,  {Carollo}
  C.~M.,  {Magee} D.,    {Gonzalez} V.,  2011, \apj, 737, 90

\bibitem[\protect\citeauthoryear{{Caputi}, {Cirasuolo}, {Dunlop}, {McLure},
  {Farrah} \& {Almaini}}{{Caputi} et~al.}{2011}]{Caputi11}
{Caputi} K.~I.,  {Cirasuolo} M.,  {Dunlop} J.~S.,  {McLure} R.~J.,  {Farrah}
  D.,    {Almaini} O.,  2011, \mnras, 413, 162

\bibitem[\protect\citeauthoryear{{Chabrier}}{{Chabrier}}{2003}]{chabrier03}
{Chabrier} G.,  2003, \pasp, 115, 763

\bibitem[\protect\citeauthoryear{{Choi} \& {Nagamine}}{{Choi} \&
  {Nagamine}}{2011}]{choina11}
{Choi} J.-H.,  {Nagamine} K.,  2011, \mnras, 410, 2579

\bibitem[\protect\citeauthoryear{{Daddi}, {Dannerbauer}, {Stern}, {Dickinson},
  {Morrison}, {Elbaz}, {Giavalisco}, {Mancini}, {Pope} \& {Spinrad}}{{Daddi}
  et~al.}{2009}]{Daddi2009}
{Daddi} E.,  {Dannerbauer} H.,  {Stern} D.,  {Dickinson} M.,  {Morrison} G.,
  {Elbaz} D.,  {Giavalisco} M.,  {Mancini} C.,  {Pope} A.,    {Spinrad} H.,
  2009, \apj, 694, 1517

\bibitem[\protect\citeauthoryear{{Daddi}, {Dickinson}, {Morrison}, {Chary},
  {Cimatti}, {Elbaz}, {Frayer}, {Renzini}, {Pope}, {Alexander}, {Bauer},
  {Giavalisco}, {Huynh}, {Kurk} \& {Mignoli}}{{Daddi} et~al.}{2007}]{Daddi2007}
{Daddi} E.,  {Dickinson} M.,  {Morrison} G.,  {Chary} R.,  {Cimatti} A.,
  {Elbaz} D.,  {Frayer} D.,  {Renzini} A.,  {Pope} A.,  {Alexander} D.~M.,
  {Bauer} F.~E.,  {Giavalisco} M.,  {Huynh} M.,  {Kurk} J.,    {Mignoli} M.,
  2007, \apj, 670, 156

\bibitem[\protect\citeauthoryear{{Dav{\'e}}}{{Dav{\'e}}}{2008}]{Dave08}
{Dav{\'e}} R.,  2008, \mnras, 385, 147

\bibitem[\protect\citeauthoryear{{Dav{\'e}}, {Oppenheimer} \&
  {Finlator}}{{Dav{\'e}} et~al.}{2011}]{Dave2011}
{Dav{\'e}} R.,  {Oppenheimer} B.~D.,    {Finlator} K.,  2011, \mnras, 415, 11

\bibitem[\protect\citeauthoryear{{Dayal} \& {Ferrara}}{{Dayal} \&
  {Ferrara}}{2012}]{DayalFerrara2012}
{Dayal} P.,  {Ferrara} A.,  2012, \mnras, 421, 2568

\bibitem[\protect\citeauthoryear{{de Barros}, {Schaerer} \& {Stark}}{{de
  Barros} et~al.}{2014}]{deBarros}
{de Barros} S.,  {Schaerer} D.,    {Stark} D.,  2014, arxiv:1207.3663

\bibitem[\protect\citeauthoryear{{Dolag}, {Borgani}, {Murante} \&
  {Springel}}{{Dolag} et~al.}{2009}]{Dolag2009}
{Dolag} K.,  {Borgani} S.,  {Murante} G.,    {Springel} V.,  2009, \mnras, 399,
  497

\bibitem[\protect\citeauthoryear{{Drory} \& {Alvarez}}{{Drory} \&
  {Alvarez}}{2008}]{Drory08}
{Drory} N.,  {Alvarez} M.,  2008, \apj, 680, 41

\bibitem[\protect\citeauthoryear{{Drory}, {Salvato}, {Gabasch}, {Bender},
  {Hopp}, {Feulner} \& {Pannella}}{{Drory} et~al.}{2005}]{Drory2005}
{Drory} N.,  {Salvato} M.,  {Gabasch} A.,  {Bender} R.,  {Hopp} U.,  {Feulner}
  G.,    {Pannella} M.,  2005, \apjl, 619, L131

\bibitem[\protect\citeauthoryear{{Duncan}, {Conselice}, {Mortlock}, {Hartley}
  \& others.}{{Duncan} et~al.}{2014}]{Duncun2014}
{Duncan} K.,  {Conselice} C.~J.,  {Mortlock} A.,  {Hartley} W.~G.,    others.
  2014, \mnras, 444, 2960

\bibitem[\protect\citeauthoryear{{Dunlop}, {McLure}, {Robertson}, {Ellis},
  {Stark}, {Cirasuolo} \& {de Ravel}}{{Dunlop} et~al.}{2012}]{Dunlop12}
{Dunlop} J.~S.,  {McLure} R.~J.,  {Robertson} B.~E.,  {Ellis} R.~S.,  {Stark}
  D.~P.,  {Cirasuolo} M.,    {de Ravel} L.,  2012, \mnras, 420, 901

\bibitem[\protect\citeauthoryear{{Dutton}, {van den Bosch} \& {Dekel}}{{Dutton}
  et~al.}{2010}]{Dutton10}
{Dutton} A.~A.,  {van den Bosch} F.~C.,    {Dekel} A.,  2010, \mnras, 405, 1690

\bibitem[\protect\citeauthoryear{{Elbaz}, {Daddi}, {Le Borgne}, {Dickinson},
  {Alexander}, {Chary}, {Starck}, {Brandt}, {Kitzbichler}, {MacDonald},
  {Nonino}, {Popesso}, {Stern} \& {Vanzella}}{{Elbaz} et~al.}{2007}]{Elbaz07}
{Elbaz} D.,  {Daddi} E.,  {Le Borgne} D.,  {Dickinson} M.,  {Alexander} D.~M.,
  {Chary} R.-R.,  {Starck} J.-L.,  {Brandt} W.~N.,  {Kitzbichler} M.,
  {MacDonald} E.,  {Nonino} M.,  {Popesso} P.,  {Stern} D.,    {Vanzella} E.,
  2007, \aap, 468, 33

\bibitem[\protect\citeauthoryear{{Ferland}, {Porter}, {van Hoof}, {Williams},
  {Abel}, {Lykins}, {Shaw}, {Henney} \& {Stancil}}{{Ferland}
  et~al.}{2013}]{ferland13}
{Ferland} G.~J.,  {Porter} R.~L.,  {van Hoof} P.~A.~M.,  {Williams} R.~J.~R.,
  {Abel} N.~P.,  {Lykins} M.~L.,  {Shaw} G.,  {Henney} W.~J.,    {Stancil}
  P.~C.,  2013, arxiv:1302.448

\bibitem[\protect\citeauthoryear{{Feulner}, {Gabasch}, {Salvato}, {Drory},
  {Hopp} \& {Bender}}{{Feulner} et~al.}{2005}]{feulner2005}
{Feulner} G.,  {Gabasch} A.,  {Salvato} M.,  {Drory} N.,  {Hopp} U.,
  {Bender} R.,  2005, \apjl, 633, L9

\bibitem[\protect\citeauthoryear{{Finkelstein}, {Cohen}, {Moustakas},
  {Malhotra}, {Rhoads} \& {Papovich}}{{Finkelstein}
  et~al.}{2011}]{Finkelstein11}
{Finkelstein} S.~L.,  {Cohen} S.~H.,  {Moustakas} J.,  {Malhotra} S.,  {Rhoads}
  J.~E.,    {Papovich} C.,  2011, \apj, 733, 117

\bibitem[\protect\citeauthoryear{{Finlator} \& {Dav{\'e}}}{{Finlator} \&
  {Dav{\'e}}}{2008}]{Fin2008}
{Finlator} K.,  {Dav{\'e}} R.,  2008, \mnras, 385, 2181

\bibitem[\protect\citeauthoryear{{Finlator}, {Dav{\'e}} \&
  {{\"O}zel}}{{Finlator} et~al.}{2011}]{Finlator11}
{Finlator} K.,  {Dav{\'e}} R.,    {{\"O}zel} F.,  2011, \apj, 743, 169

\bibitem[\protect\citeauthoryear{{Finlator}, {Oppenheimer} \&
  {Dav{\'e}}}{{Finlator} et~al.}{2011}]{Finlator11b}
{Finlator} K.,  {Oppenheimer} B.~D.,    {Dav{\'e}} R.,  2011, \mnras, 410, 1703

\bibitem[\protect\citeauthoryear{{Gonz{\'a}lez}, {Bouwens}, {Labb{\'e}},
  {Illingworth}, {Oesch}, {Franx} \& {Magee}}{{Gonz{\'a}lez}
  et~al.}{2012}]{Gonzalez12}
{Gonz{\'a}lez} V.,  {Bouwens} R.~J.,  {Labb{\'e}} I.,  {Illingworth} G.,
  {Oesch} P.,  {Franx} M.,    {Magee} D.,  2012, \apj, 755, 148

\bibitem[\protect\citeauthoryear{{Gonz{\'a}lez}, {Labb{\'e}}, {Bouwens},
  {Illingworth}, {Franx} \& {Kriek}}{{Gonz{\'a}lez} et~al.}{2011}]{Gonzalez11}
{Gonz{\'a}lez} V.,  {Labb{\'e}} I.,  {Bouwens} R.~J.,  {Illingworth} G.,
  {Franx} M.,    {Kriek} M.,  2011, \apjl, 735, L34

\bibitem[\protect\citeauthoryear{{Gonz{\'a}lez}, {Labb{\'e}}, {Bouwens},
  {Illingworth}, {Franx}, {Kriek} \& {Brammer}}{{Gonz{\'a}lez}
  et~al.}{2010}]{Gonzalez10}
{Gonz{\'a}lez} V.,  {Labb{\'e}} I.,  {Bouwens} R.~J.,  {Illingworth} G.,
  {Franx} M.,  {Kriek} M.,    {Brammer} G.~B.,  2010, \apj, 713, 115

\bibitem[\protect\citeauthoryear{{Haardt} \& {Madau}}{{Haardt} \&
  {Madau}}{2001}]{haardtmadau01}
{Haardt} F.,  {Madau} P.,  2001, in {Neumann} D.~M.,  {Tran} J.~T.~V.,  eds,
  Clusters of Galaxies and the High Redshift Universe Observed in X-rays
  {Modelling the UV/X-ray cosmic background with CUBA}

\bibitem[\protect\citeauthoryear{{Haas}, {Schaye}, {Booth}, {Dalla Vecchia},
  {Springel}, {Theuns} \& {Wiersma}}{{Haas} et~al.}{2013a}]{Haas2013}
{Haas} M.~R.,  {Schaye} J.,  {Booth} C.~M.,  {Dalla Vecchia} C.,  {Springel}
  V.,  {Theuns} T.,    {Wiersma} R.~P.~C.,  2013a, \mnras, 435, 2931

\bibitem[\protect\citeauthoryear{{Haas}, {Schaye}, {Booth}, {Dalla Vecchia},
  {Springel}, {Theuns} \& {Wiersma}}{{Haas} et~al.}{2013b}]{Haas2013b}
{Haas} M.~R.,  {Schaye} J.,  {Booth} C.~M.,  {Dalla Vecchia} C.,  {Springel}
  V.,  {Theuns} T.,    {Wiersma} R.~P.~C.,  2013b, \mnras, 435, 2955

\bibitem[\protect\citeauthoryear{{Heinis}, {Buat}, {B{\'e}thermin}, {Bock},
  {Burgarella}, {Conley}, {Cooray}, {Farrah}, {Ilbert}, {Magdis}, {Marsden},
  {Oliver}, {Rigopoulou}, {Roehlly}, {Schulz}, {Symeonidis}, {Viero}, {Xu} \&
  {Zemcov}}{{Heinis} et~al.}{2014}]{Heinis2014}
{Heinis} S.,  {Buat} V.,  {B{\'e}thermin} M.,  {Bock} J.,  {Burgarella} D.,
  {Conley} A.,  {Cooray} A.,  {Farrah} D.,  {Ilbert} O.,  {Magdis} G.,
  {Marsden} G.,  {Oliver} S.~J.,  {Rigopoulou} D.,  {Roehlly} Y.,  {Schulz} B.,
   {Symeonidis} M.,  {Viero} M.,  {Xu} C.~K.,    {Zemcov} M.,  2014, \mnras,
  437, 1268

\bibitem[\protect\citeauthoryear{{Jaacks}, {Choi}, {Nagamine}, {Thompson} \&
  {Varghese}}{{Jaacks} et~al.}{2012}]{jaacks12}
{Jaacks} J.,  {Choi} J.-H.,  {Nagamine} K.,  {Thompson} R.,    {Varghese} S.,
  2012, \mnras, 420, 1606

\bibitem[\protect\citeauthoryear{{Kannan}, {Stinson}, {Macci{\`o}}, {Brook},
  {Weinmann}, {Wadsley} \& {Couchman}}{{Kannan} et~al.}{2014}]{Kannan2014}
{Kannan} R.,  {Stinson} G.~S.,  {Macci{\`o}} A.~V.,  {Brook} C.,  {Weinmann}
  S.~M.,  {Wadsley} J.,    {Couchman} H.~M.~P.,  2014, \mnras, 437, 3529

\bibitem[\protect\citeauthoryear{{Karim}, {Schinnerer},
  {Mart{\'{\i}}nez-Sansigre} \& others.}{{Karim} et~al.}{2011}]{Karim2011}
{Karim} A.,  {Schinnerer} E.,  {Mart{\'{\i}}nez-Sansigre} A.,    others. 2011,
  \apj, 730, 61

\bibitem[\protect\citeauthoryear{{Kennicutt}
  Jr.}{{Kennicutt}}{1998}]{kennicutt1998}
{Kennicutt} Jr. R.~C.,  1998, \araa, 36, 189

\bibitem[\protect\citeauthoryear{{Komatsu} et~al.,}{{Komatsu}
  et~al.}{2011}]{komatsu11}
{Komatsu} E.,  et~al., 2011, \apjs, 192, 18

\bibitem[\protect\citeauthoryear{{Kroupa}, {Tout} \& {Gilmore}}{{Kroupa}
  et~al.}{1993}]{kroupa93}
{Kroupa} P.,  {Tout} C.~A.,    {Gilmore} G.,  1993, \mnras, 262, 545

\bibitem[\protect\citeauthoryear{{Labb{\'e}}, {Franx}, {Rudnick}, {Schreiber},
  {Rix}, {Moorwood}, {van Dokkum}, {van der Werf}, {R{\"o}ttgering}, {van
  Starkenburg}, {van der Wel}, {Kuijken} \& {Daddi}}{{Labb{\'e}}
  et~al.}{2003}]{Labbe2003}
{Labb{\'e}} I.,  {Franx} M.,  {Rudnick} G.,  {Schreiber} N.~M.~F.,  {Rix}
  H.-W.,  {Moorwood} A.,  {van Dokkum} P.~G.,  {van der Werf} P.,
  {R{\"o}ttgering} H.,  {van Starkenburg} L.,  {van der Wel} A.,  {Kuijken} K.,
     {Daddi} E.,  2003, \aj, 125, 1107

\bibitem[\protect\citeauthoryear{{Labb{\'e}}, {Gonz{\'a}lez}, {Bouwens},
  {Illingworth}, {Oesch}, {van Dokkum}, {Carollo}, {Franx}, {Stiavelli},
  {Trenti}, {Magee} \& {Kriek}}{{Labb{\'e}} et~al.}{2010}]{Labbe10}
{Labb{\'e}} I.,  {Gonz{\'a}lez} V.,  {Bouwens} R.~J.,  {Illingworth} G.~D.,
  {Oesch} P.~A.,  {van Dokkum} P.~G.,  {Carollo} C.~M.,  {Franx} M.,
  {Stiavelli} M.,  {Trenti} M.,  {Magee} D.,    {Kriek} M.,  2010, \apjl, 708,
  L26

\bibitem[\protect\citeauthoryear{{Lee}, {Dey}, {Reddy}, {Brown}, {Gonzalez},
  {Jannuzi}, {Cooper}, {Fan}, {Bian}, {Glikman}, {Stern}, {Brodwin} \&
  {Cooray}}{{Lee} et~al.}{2011}]{Lee11}
{Lee} K.-S.,  {Dey} A.,  {Reddy} N.,  {Brown} M.~J.~I.,  {Gonzalez} A.~H.,
  {Jannuzi} B.~T.,  {Cooper} M.~C.,  {Fan} X.,  {Bian} F.,  {Glikman} E.,
  {Stern} D.,  {Brodwin} M.,    {Cooray} A.,  2011, \apj, 733, 99

\bibitem[\protect\citeauthoryear{{Lee}, {Ferguson}, {Wiklind}, {Dahlen},
  {Dickinson}, {Giavalisco}, {Grogin}, {Papovich}, {Messias}, {Guo} \&
  {Lin}}{{Lee} et~al.}{2012}]{Lee2012}
{Lee} K.-S.,  {Ferguson} H.~C.,  {Wiklind} T.,  {Dahlen} T.,  {Dickinson}
  M.~E.,  {Giavalisco} M.,  {Grogin} N.,  {Papovich} C.,  {Messias} H.,  {Guo}
  Y.,    {Lin} L.,  2012, \apj, 752, 66

\bibitem[\protect\citeauthoryear{{Lo Faro}, {Monaco}, {Vanzella}, {Fontanot},
  {Silva} \& {Cristiani}}{{Lo Faro} et~al.}{2009}]{Lofaro09}
{Lo Faro} B.,  {Monaco} P.,  {Vanzella} E.,  {Fontanot} F.,  {Silva} L.,
  {Cristiani} S.,  2009, \mnras, 399, 827

\bibitem[\protect\citeauthoryear{{Madau}, {Pozzetti} \& {Dickinson}}{{Madau}
  et~al.}{1998}]{madau1998}
{Madau} P.,  {Pozzetti} L.,    {Dickinson} M.,  1998, \apj, 498, 106

\bibitem[\protect\citeauthoryear{{Magdis}, {Rigopoulou}, {Huang} \&
  {Fazio}}{{Magdis} et~al.}{2010}]{Magdis10}
{Magdis} G.~E.,  {Rigopoulou} D.,  {Huang} J.-S.,    {Fazio} G.~G.,  2010,
  \mnras, 401, 1521

\bibitem[\protect\citeauthoryear{{Magorrian}, {Tremaine}, {Richstone},
  {Bender}, {Bower}, {Dressler}, {Faber}, {Gebhardt}, {Green}, {Grillmair},
  {Kormendy} \& {Lauer}}{{Magorrian} et~al.}{1998}]{magorrian1998}
{Magorrian} J.,  {Tremaine} S.,  {Richstone} D.,  {Bender} R.,  {Bower} G.,
  {Dressler} A.,  {Faber} S.~M.,  {Gebhardt} K.,  {Green} R.,  {Grillmair} C.,
  {Kormendy} J.,    {Lauer} T.,  1998, \aj, 115, 2285

\bibitem[\protect\citeauthoryear{{Marchesini}, {van Dokkum}, {F{\"o}rster
  Schreiber}, {Franx}, {Labb{\'e}} \& {Wuyts}}{{Marchesini}
  et~al.}{2009}]{Marchesini09}
{Marchesini} D.,  {van Dokkum} P.~G.,  {F{\"o}rster Schreiber} N.~M.,  {Franx}
  M.,  {Labb{\'e}} I.,    {Wuyts} S.,  2009, \apj, 701, 1765

\bibitem[\protect\citeauthoryear{{Martin}}{{Martin}}{2005}]{Martin05}
{Martin} C.~L.,  2005, \apj, 621, 227

\bibitem[\protect\citeauthoryear{{McLure}, {Dunlop}, {de Ravel}, {Cirasuolo},
  {Ellis}, {Schenker}, {Robertson}, {Koekemoer}, {Stark} \& {Bowler}}{{McLure}
  et~al.}{2011}]{MClure2011}
{McLure} R.~J.,  {Dunlop} J.~S.,  {de Ravel} L.,  {Cirasuolo} M.,  {Ellis}
  R.~S.,  {Schenker} M.,  {Robertson} B.~E.,  {Koekemoer} A.~M.,  {Stark}
  D.~P.,    {Bowler} R.~A.~A.,  2011, \mnras, 418, 2074

\bibitem[\protect\citeauthoryear{{Menci}, {Fontana}, {Giallongo}, {Grazian} \&
  {Salimbeni}}{{Menci} et~al.}{2006}]{Menci06}
{Menci} N.,  {Fontana} A.,  {Giallongo} E.,  {Grazian} A.,    {Salimbeni} S.,
  2006, \apj, 647, 753

\bibitem[\protect\citeauthoryear{{Meurer}, {Heckman} \& {Calzetti}}{{Meurer}
  et~al.}{1999}]{meurer1999}
{Meurer} G.~R.,  {Heckman} T.~M.,    {Calzetti} D.,  1999, \apj, 521, 64

\bibitem[\protect\citeauthoryear{{Monaco}, {Fontanot} \& {Taffoni}}{{Monaco}
  et~al.}{2007}]{Monaco07}
{Monaco} P.,  {Fontanot} F.,    {Taffoni} G.,  2007, \mnras, 375, 1189

\bibitem[\protect\citeauthoryear{{Noeske} et~al.,}{{Noeske}
  et~al.}{2007}]{Noeske2007}
{Noeske} K.~G.,  et~al., 2007, \apjl, 660, L43

\bibitem[\protect\citeauthoryear{{Oppenheimer} \& {Dav{\'e}}}{{Oppenheimer} \&
  {Dav{\'e}}}{2006}]{oppe06}
{Oppenheimer} B.~D.,  {Dav{\'e}} R.,  2006, \mnras, 373, 1265

\bibitem[\protect\citeauthoryear{{Papovich}, {Finkelstein}, {Ferguson}, {Lotz}
  \& {Giavalisco}}{{Papovich} et~al.}{2011}]{Papov11}
{Papovich} C.,  {Finkelstein} S.~L.,  {Ferguson} H.~C.,  {Lotz} J.~M.,
  {Giavalisco} M.,  2011, \mnras, 412, 1123

\bibitem[\protect\citeauthoryear{{Percival} et~al.,}{{Percival}
  et~al.}{2010}]{percival10}
{Percival} W.~J.,  et~al., 2010, \mnras, 401, 2148

\bibitem[\protect\citeauthoryear{{Planck Collaboration}, {Ade}, {Aghanim},
  {Armitage-Caplan}, {Arnaud}, {Ashdown}, {Atrio-Barandela}, {Aumont},
  {Baccigalupi}, {Banday} \& et al.}{{Planck Collaboration}
  et~al.}{2013}]{Planck13}
{Planck Collaboration} {Ade} P.~A.~R.,  {Aghanim} N.,  {Armitage-Caplan} C.,
  {Arnaud} M.,  {Ashdown} M.,  {Atrio-Barandela} F.,  {Aumont} J.,
  {Baccigalupi} C.,  {Banday} A.~J.,    et al. 2013, arxiv:1303.5076

\bibitem[\protect\citeauthoryear{{Puchwein} \& {Springel}}{{Puchwein} \&
  {Springel}}{2013}]{PuchweinSpri12}
{Puchwein} E.,  {Springel} V.,  2013, \mnras, 428, 2966

\bibitem[\protect\citeauthoryear{{Reddy}, {Pettini}, {Steidel}, {Shapley},
  {Erb} \& {Law}}{{Reddy} et~al.}{2012}]{reddy2012}
{Reddy} N.~A.,  {Pettini} M.,  {Steidel} C.~C.,  {Shapley} A.~E.,  {Erb} D.~K.,
     {Law} D.~R.,  2012, \apj, 754, 25

\bibitem[\protect\citeauthoryear{{Riess}, {Macri}, {Casertano}, {Sosey},
  {Lampeitl}, {Ferguson}, {Filippenko}, {Jha}, {Li}, {Chornock} \&
  {Sarkar}}{{Riess} et~al.}{2009}]{riess09}
{Riess} A.~G.,  {Macri} L.,  {Casertano} S.,  {Sosey} M.,  {Lampeitl} H.,
  {Ferguson} H.~C.,  {Filippenko} A.~V.,  {Jha} S.~W.,  {Li} W.,  {Chornock}
  R.,    {Sarkar} D.,  2009, \apj, 699, 539

\bibitem[\protect\citeauthoryear{{Salim}, {Rich}, {Charlot}, {Brinchmann}
  et~al.,}{{Salim} et~al.}{2007}]{Salim2007}
{Salim} S.,  {Rich} R.~M.,  {Charlot} S.,  {Brinchmann} J.,    et~al., 2007,
  \apjs, 173, 267

\bibitem[\protect\citeauthoryear{{Salmon}, {Papovich}, {Finkelstein} \&
  others.}{{Salmon} et~al.}{2014}]{Salmon2014}
{Salmon} B.,  {Papovich} C.,  {Finkelstein} S.~L.,    others. 2014, ArXiv
  e-prints

\bibitem[\protect\citeauthoryear{{Salpeter}}{{Salpeter}}{1955}]{salpeter55}
{Salpeter} E.~E.,  1955, \apj, 121, 161

\bibitem[\protect\citeauthoryear{{Santini}, {Fontana}, {Grazian}, {Salimbeni},
  {Fontanot}, {Paris}, {Boutsia}, {Castellano}, {Fiore}, {Gallozzi},
  {Giallongo}, {Koekemoer}, {Menci}, {Pentericci} \& {Somerville}}{{Santini}
  et~al.}{2012}]{Santini12}
{Santini} P.,  {Fontana} A.,  {Grazian} A.,  {Salimbeni} S.,  {Fontanot} F.,
  {Paris} D.,  {Boutsia} K.,  {Castellano} M.,  {Fiore} F.,  {Gallozzi} S.,
  {Giallongo} E.,  {Koekemoer} A.~M.,  {Menci} N.,  {Pentericci} L.,
  {Somerville} R.~S.,  2012, \aap, 538, A33

\bibitem[\protect\citeauthoryear{{Sawicki}}{{Sawicki}}{2012}]{Sawiki}
{Sawicki} M.,  2012, \mnras, 421, 2187

\bibitem[\protect\citeauthoryear{{Schaye}, {Dalla Vecchia}, {Booth}, {Wiersma},
  {Theuns}, {Haas}, {Bertone}, {Duffy}, {McCarthy} \& {van de Voort}}{{Schaye}
  et~al.}{2010}]{schaye10}
{Schaye} J.,  {Dalla Vecchia} C.,  {Booth} C.~M.,  {Wiersma} R.~P.~C.,
  {Theuns} T.,  {Haas} M.~R.,  {Bertone} S.,  {Duffy} A.~R.,  {McCarthy} I.~G.,
     {van de Voort} F.,  2010, \mnras, 402, 1536

\bibitem[\protect\citeauthoryear{{Shapley}, {Steidel}, {Erb}, {Reddy},
  {Adelberger}, {Pettini}, {Barmby} \& {Huang}}{{Shapley}
  et~al.}{2005}]{Shapley05}
{Shapley} A.~E.,  {Steidel} C.~C.,  {Erb} D.~K.,  {Reddy} N.~A.,  {Adelberger}
  K.~L.,  {Pettini} M.,  {Barmby} P.,    {Huang} J.,  2005, \apj, 626, 698

\bibitem[\protect\citeauthoryear{{Smit}, {Bouwens}, {Franx}, {Illingworth},
  {Labb{\'e}}, {Oesch} \& {van Dokkum}}{{Smit} et~al.}{2012}]{smit12}
{Smit} R.,  {Bouwens} R.~J.,  {Franx} M.,  {Illingworth} G.~D.,  {Labb{\'e}}
  I.,  {Oesch} P.~A.,    {van Dokkum} P.~G.,  2012, \apj, 756, 14

\bibitem[\protect\citeauthoryear{{Smit}, {Bouwens}, {Labbe}, {Zheng}
  et~al.,}{{Smit} et~al.}{2013}]{Smit2013}
{Smit} R.,  {Bouwens} R.~J.,  {Labbe} I.,  {Zheng} W.,    et~al., 2013,
  arxiv:1307.5847

\bibitem[\protect\citeauthoryear{{Somerville}, {Gilmore}, {Primack} \&
  {Dom{\'{\i}}nguez}}{{Somerville} et~al.}{2012}]{Somerville12}
{Somerville} R.~S.,  {Gilmore} R.~C.,  {Primack} J.~R.,    {Dom{\'{\i}}nguez}
  A.,  2012, \mnras, 423, 1992

\bibitem[\protect\citeauthoryear{{Springel}}{{Springel}}{2005}]{springel2005}
{Springel} V.,  2005, \mnras, 364, 1105

\bibitem[\protect\citeauthoryear{{Springel} \& {Hernquist}}{{Springel} \&
  {Hernquist}}{2003}]{springel2003}
{Springel} V.,  {Hernquist} L.,  2003, \mnras, 339, 289

\bibitem[\protect\citeauthoryear{{Stark}, {Ellis}, {Bunker}, {Bundy},
  {Targett}, {Benson} \& {Lacy}}{{Stark} et~al.}{2009}]{stark2009}
{Stark} D.~P.,  {Ellis} R.~S.,  {Bunker} A.,  {Bundy} K.,  {Targett} T.,
  {Benson} A.,    {Lacy} M.,  2009, \apj, 697, 1493

\bibitem[\protect\citeauthoryear{{Stark}, {Schenker}, {Ellis}, {Robertson},
  {McLure} \& {Dunlop}}{{Stark} et~al.}{2013}]{StarkSchen}
{Stark} D.~P.,  {Schenker} M.~A.,  {Ellis} R.,  {Robertson} B.,  {McLure} R.,
   {Dunlop} J.,  2013, \apj, 763, 129

\bibitem[\protect\citeauthoryear{{Tescari}, {Katsianis}, {Wyithe}, {Dolag},
  {Tornatore}, {Barai}, {Viel} \& {Borgani}}{{Tescari}
  et~al.}{2014}]{TescariKaW2013}
{Tescari} E.,  {Katsianis} A.,  {Wyithe} J.~S.~B.,  {Dolag} K.,  {Tornatore}
  L.,  {Barai} P.,  {Viel} M.,    {Borgani} S.,  2014, \mnras, 438, 3490

\bibitem[\protect\citeauthoryear{{Tescari}, {Viel}, {D'Odorico}, {Cristiani},
  {Calura}, {Borgani} \& {Tornatore}}{{Tescari} et~al.}{2011}]{tex11}
{Tescari} E.,  {Viel} M.,  {D'Odorico} V.,  {Cristiani} S.,  {Calura} F.,
  {Borgani} S.,    {Tornatore} L.,  2011, \mnras, 411, 826

\bibitem[\protect\citeauthoryear{{Tescari}, {Viel}, {Tornatore} \&
  {Borgani}}{{Tescari} et~al.}{2009}]{tescari09}
{Tescari} E.,  {Viel} M.,  {Tornatore} L.,    {Borgani} S.,  2009, \mnras, 397,
  411

\bibitem[\protect\citeauthoryear{{Tornatore}, {Borgani}, {Dolag} \&
  {Matteucci}}{{Tornatore} et~al.}{2007}]{T07}
{Tornatore} L.,  {Borgani} S.,  {Dolag} K.,    {Matteucci} F.,  2007, \mnras,
  382, 1050

\bibitem[\protect\citeauthoryear{{Vogelsberger}, {Genel}, {Sijacki}, {Torrey},
  {Springel} \& {Hernquist}}{{Vogelsberger} et~al.}{2013}]{Vogel13}
{Vogelsberger} M.,  {Genel} S.,  {Sijacki} D.,  {Torrey} P.,  {Springel} V.,
  {Hernquist} L.,  2013, ArXiv e-prints: 1305.2913

\bibitem[\protect\citeauthoryear{{Wang}, {De Lucia}, {Kitzbichler} \&
  {White}}{{Wang} et~al.}{2008}]{Wang08}
{Wang} J.,  {De Lucia} G.,  {Kitzbichler} M.~G.,    {White} S.~D.~M.,  2008,
  \mnras, 384, 1301

\bibitem[\protect\citeauthoryear{{Weinmann}, {Neistein} \& {Dekel}}{{Weinmann}
  et~al.}{2011}]{Weinmann2011}
{Weinmann} S.~M.,  {Neistein} E.,    {Dekel} A.,  2011, \mnras, 417, 2737

\bibitem[\protect\citeauthoryear{{Whitaker}, {van Dokkum}, {Brammer} \&
  {Franx}}{{Whitaker} et~al.}{2012}]{Whitaker2012}
{Whitaker} K.~E.,  {van Dokkum} P.~G.,  {Brammer} G.,    {Franx} M.,  2012,
  \apjl, 754, L29

\bibitem[\protect\citeauthoryear{{Wiersma}, {Schaye} \& {Smith}}{{Wiersma}
  et~al.}{2009}]{wiersma09}
{Wiersma} R.~P.~C.,  {Schaye} J.,    {Smith} B.~D.,  2009, \mnras, 393, 99

\bibitem[\protect\citeauthoryear{{Wilkins}, {Di Matteo}, {Croft}, {Khandai},
  {Feng}, {Bunker} \& {Coulton}}{{Wilkins} et~al.}{2013}]{Wilkins13}
{Wilkins} S.~M.,  {Di Matteo} T.,  {Croft} R.,  {Khandai} N.,  {Feng} Y.,
  {Bunker} A.,    {Coulton} W.,  2013, \mnras, 429, 2098

\bibitem[\protect\citeauthoryear{{Wyithe}, {Loeb} \& {Oesch}}{{Wyithe}
  et~al.}{2013}]{wyithe13}
{Wyithe} S.,  {Loeb} A.,    {Oesch} P.,  2013, arxiv:1308.2030

\end{thebibliography}

\section*{Appendix A: Box size and resolution tests}
\label{appendix_a}

In this appendix we perform box size and resolution tests, in order to check the convergence of the results from our simulations. As stated in \citet{TescariKaW2013}, in our simulations the smaller the box size of a run, the higher its mass/spatial resolution. However, the box size sets an upper limit on the mass of the halos that can be formed in the simulated volume. Therefore, for our simulations, higher resolution means also poorer statistics at the high mass end of the halo mass function.

In the top 4 panels of Fig. \ref{fig:SFR-mass12} we compare the median lines of the SFR$-{\rm M}_{\rm FoF}$ density plots for runs with box size equal to $L=24$ Mpc/$h$, $L=18$ Mpc/$h$ and $L=12$ Mpc/$h$. Four simulations are considered: two of them in the late AGN + weak Winds setup (\textit{Ch24\tu lA\tu wW} - orange triple dot-dashed line and \textit{Ch18\tu lA\tu wW} - blue dot-dashed line) and the other two in the early AGN + strong Winds setup (\textit{Ch24\tu eA\tu sW} - red triple dot-dashed line and \textit{Ch12\tu eA\tu sW} - black dotted line). At $z\ge5.9$, all the simulations are virtually indistinguishable inside the observational limits in the range of SFR of \citet[][blue dot-dashed horizontal lines]{smit12}. The convergence of the results is slightly worse at redshift $z=5$, but at $z=3.8$ simulations with the same configuration and different box size/resolution are convergent to $\sim0.1$ dex. As expected, the run with box size $L=12$ Mpc/$h$ cannot sample halos with total mass ${\rm M_{FoF}}$ $\apprge10^{11.5}$ M$_{\odot}/h$ due to its small simulated volume.

In the bottom 4 panels of Fig. \ref{fig:SFR-mass12} we test the SFR$-{\rm M}_{\star}$ relation using the same simulations. In this case there is a very good agreement between the different runs at all redshifts considered. For comparison, we overplotted the observations of \citet[][observed frame I-band selected sample - green dashed + dotted lines]{Drory08}, \citet[][droup-outs selection - blue triple dot-dashed lines + light blue shaded regions]{bouwens2012} and \citet[][droup-outs selection - nebular emission - red filled circles]{deBarros}.

In Fig. \ref{fig:SMFres} we plot the galaxy stellar mass functions for the four runs used before and we compare with the observational results of \citet[][green filled circles with error bars]{Marchesini09}, \citet[][blue filled squares with error bars]{Gonzalez11} and \citet[][magenta dashed line]{Santini12}. The effect of poorer statistics at high stellar masses for the \textit{Ch12\tu eA\tu sW} run is clearly visible. However, in each panel our simulations show a good convergence down to the observational point of \citet{Gonzalez11} at $\log ({\rm M}{_\star}/{\rm M}_{\odot})=8.25$. Below this limit, incompleteness effects, due to the SFR/mass cut we adopted in our simulations, start to be important (the same feature is visible for the runs plotted in Figs. \ref{fig:aSMFa}, \ref{fig:aSMFaclear} and \ref{fig:aSMFb}).

To summarise, the SFR$-{\rm M}_{\star}$ relation converges at $4\le z\le7$, while the SFR$-{\rm M}_{\rm FoF}$ relation shows a slightly worse convergence at $z=5.0$. Finally, at all redshifts considered the GSMF converges down to $\log ({\rm M}{_\star}/{\rm M}_{\odot})=8.25$.

\label{lastpage}
\end{document}